\documentclass{PoS}
\usepackage{amsmath}
\usepackage{amssymb}
\usepackage{dsfont}
\usepackage{bbm}
\DeclareMathAlphabet{\mathcal}{OMS}{cmsy}{m}{n}
\usepackage[utf8]{inputenc}

\let\OLDthebibliography\thebibliography
\renewcommand\thebibliography[1]{
  \OLDthebibliography{#1}
  \setlength{\parskip}{0pt}
  \setlength{\itemsep}{0pt plus 0.3ex}
}

\title{Leading isospin-breaking effects in the hadronic vacuum polarisation with open boundaries}

\ShortTitle{Leading isospin-breaking effects in the hadronic vacuum polarisation with open boundaries}

\author{\speaker{Andreas~Risch}$^{a}$, Hartmut~Wittig$^{a,b}$\\
        $^{a}$ PRISMA Cluster of Excellence and Institut f{\"u}r Kernphysik, University of Mainz, Germany\\
        $^{b}$ Helmholtz Institute Mainz, University of Mainz, Germany\\
        E-mail: \email{andreas.risch@uni-mainz.de}, \email{hartmut.wittig@uni-mainz.de}}

\abstract{We discuss leading isospin-breaking effects in the hadronic vacuum polarisation required for the investigation of the hadronic contribution to $(g-2)_{\mu}$. The calculation proceeds by expanding the relevant correlation functions around the isosymmetric limit. Isosymmetric observables are evaluated on gauge ensembles with $N_{f}=2+1$, $O(a)$ improved Wilson fermions and open boundary conditions generated by the CLS effort. Particular emphasis is placed on the relevant quark-disconnected diagrams required for a complete treatment of leading isospin-breaking effects in the valence quark sector. We provide a detailed discussion of the renormalisation of the vector current in QCD+QED taking operator mixing into account.}

\FullConference{37th International Symposium on Lattice Field Theory - Lattice2019\\
		16-22 June 2019\\
		Wuhan, China}

\begin{document}

\section{Introduction}

In this work we continue the investigation of isospin-breaking effects making use of Coordinated Lattice Simulations (CLS) $N_{\mathrm{f}}=2+1$ QCD ensembles~\cite{Bruno:2014jqa, Bruno:2016plf} with open boundary conditions~\cite{Luscher:2011kk} first covered in~\cite{Risch:2018ozp}. For a previous account of our effort based on $N_{\mathrm{f}}=2$ QCD ensembles with (anti-)periodic boundaries conditions see~\cite{Risch:2017xxe}. We follow the ROME123 approach~\cite{deDivitiis:2011eh, deDivitiis:2013xla} which treats isospin-breaking effects perturbatively. It is desirable to calculate hadronic corrections to high-precision observables such as the anomalous magnetic moment of the muon with high accuracy as it provides intriguing hints for the possible existence of new physics. We therefore investigate the hadronic vacuum polarisation (HVP) in QCD+QED. This work is organised as follows: We recap the setup used for the perturbative treatment of isospin-breaking effects and apply it to the bare vector-vector correlation function. We further describe the renormalisation procedure of the local vector current taking operator mixing into account and determine the relevant renormalisation factors. We construct the renormalised hadronic vacuum polarisation function and discuss the status of our investigation of quark-disconnected contributions.

\section{Inclusion of perturbative isospin-breaking effects by reweighting} \label{section2}

We briefly summarise our setup for the perturbative treatment of isospin-breaking effects. For a detailed description we refer to \cite{Risch:2018ozp}. We consider the space of QCD+QED-like theories parameterised by $\varepsilon = (m_{\mathrm{u}}, m_{\mathrm{d}},  m_{\mathrm{s}},\beta, e^{2})$. For the choice $\varepsilon^{(0)} = (m_{\mathrm{u}}^{(0)}, m_{\mathrm{d}}^{(0)}, m_{\mathrm{s}}^{(0)}, \beta^{(0)}, 0)$ with $m_{\mathrm{u}}^{(0)}= m_{\mathrm{d}}^{(0)}$ we obtain QCD$_{\mathrm{iso}}$ together with a free photon field. In~\cite{Risch:2018ozp} we have shown that QCD+QED can be related to QCD$_{\mathrm{iso}}$ by reweighting via the identity
\begin{align}
\langle O[U,A,\Psi,\overline{\Psi}] \rangle &= \frac{\langle R[U] \langle O[U,A,\Psi,\overline{\Psi}] \rangle_{\mathrm{q}\gamma} \rangle_{\mathrm{eff}}^{(0)}}{\langle R[U] \rangle_{\mathrm{eff}}^{(0)}} & R[U] &= \frac{\exp(-S_{\mathrm{g}}[U])Z_{\mathrm{q}\gamma}[U]}{\exp(-S_{\mathrm{g}}^{(0)}[U])Z^{(0)}_{\mathrm{q}}[U]},
\label{eq_expectation_value_by_reweighting}
\end{align}
where $\langle \ldots \rangle_{\mathrm{eff}}^{(0)}$ is evaluated by making use of existing QCD$_{\mathrm{iso}}$ gauge configurations and $\left\langle \ldots \right\rangle_{\mathrm{q}\gamma}$ and $R[U]$ are evaluated by means of perturbation theory in $\Delta\varepsilon=\varepsilon-\varepsilon^{(0)}$ around $\varepsilon^{(0)}$. The required Feynman rules are discussed in~\cite{Risch:2018ozp}. In order to fix the expansion coefficients $\Delta\varepsilon$ we make use of a suitable hadronic renormalisation scheme~\cite{Risch:2018ozp}.

The simulation code is based on the QDP++~\cite{Edwards:2004sx} and FFTW3~\cite{FFTW05} libraries and the openQCD~\cite{Luscher:} framework. So far, we have performed simulations on one gauge ensemble (c.f. Table~\ref{table_lattice_parameters}) containing $2004$ configurations with a binning size of four. Quark-connected diagrams were estimated with $32$ stochastic $U(1)$-quark-sources at the mesonic source distributed on the time-slices $26-35$ and $60-69$ utilising the approximate translational invariance in the bulk of the lattice. The QED$_{\mathrm{L}}$ photon propagator~\cite{Risch:2018ozp} was estimated with two $Z_{2}$-photon-sources per quark-source.
\begin{table}[b]
  \centering
\begin{tabular}{|l|l|l|l|l|l|l|}
\hline
 & $\mathrm{T}/a \times (\mathrm{L}/a)^3$ & $\beta$ & $a\,[\mathrm{fm}]$ & $m_{\pi}\,[\mathrm{MeV}]$ & $m_{K}\,[\mathrm{MeV}]$ & $m_{\pi}L$ \\ \hline
 H102 & $96 \times 32^{3}$ & $3.4$ & $0.08636(98)(40)$ & $350$ & $440$ & $4.9$ \\ \hline
\end{tabular}
\caption{Parameters of CLS open boundary ensembles with $N_{\mathrm{f}}=2+1$ quark flavours of non-perturbatively O(a) improved Wilson quarks and tree-level improved L\"uscher-Weisz gauge action~\cite{Bruno:2014jqa, Bruno:2016plf}.}
\label{table_lattice_parameters}
\end{table}

\section{The bare vector-vector correlation function}

In the Mainz setup~\cite{Gerardin:2019rua, Ce:2019imp} we use both the local $V^{\gamma}_{\mathrm{l}}$ and conserved vector currents $V^{\gamma}_{\mathrm{c}}$. Employing open boundary conditions we only give the bulk part of the operators for simplicity:
\begin{align*}
V{}^{\gamma x\mu}_{\mathrm{l}} &= \overline{\Psi}{}^{x+\hat{\mu}}Q\gamma^{\mu}\Psi{}^{x} & V{}^{\gamma x\mu}_{\mathrm{c}} &= \frac{1}{2}\big(\overline{\Psi}{}^{x+\hat{\mu}}Q(\gamma^{\mu}+\mathds{1})(W^{x\mu})^{\dagger}\Psi{}^{x}+\overline{\Psi}{}^{x}Q(\gamma^{\mu}-\mathds{1})W^{x\mu}\Psi{}^{x+\hat{\mu}}\big).
\end{align*}
$W{}^{x\mu}=U{}^{x\mu}\exp(\mathrm{i}eQA{}^{x\mu})$ are the combined QCD+QED gauge links with the matrix of fractional quark charges $Q=\mathrm{diag}\big(\frac{2}{3},-\frac{1}{3},-\frac{1}{3}\big)$. As we will see later in the context of renormalisation it is convenient to decompose the electromagnetic current as $V^{\gamma}=V^{3}+\frac{1}{\sqrt{3}}V^{8}$, making use of
\begin{align}
V_{\mathrm{l}}{}^{x\mu i} &= \overline{\Psi}{}^{x}\Lambda^{i}\gamma^{\mu}\Psi{}^{x} & V_{\mathrm{c}}{}^{x\mu i} &= \frac{1}{2}\big(\overline{\Psi}{}^{x+\hat{\mu}}\Lambda^{i}(\gamma^{\mu}+\mathds{1})(W^{x\mu})^{\dagger}\Psi{}^{x}+\overline{\Psi}{}^{x}\Lambda^{i}(\gamma^{\mu}-\mathds{1})W^{x\mu}\Psi{}^{x+\hat{\mu}}\big)
\label{eq_v_currents}
\end{align}
with $\Lambda^{0} = \frac{1}{\sqrt{6}}\mathds{1}$ and $\Lambda^{i} = \frac{1}{2}\lambda^{i}$ for $i=3,8$. We investigate correlation functions of the form
\begin{align}
C^{i_{2}i_{1}}_{d_{2}\mathrm{l}}(x_{2}^{0},x_{1}^{0}) &= \frac{1}{|\Lambda_{123}|}\sum_{\vec{x}_{1}, \vec{x}_{2}}\frac{1}{3}\sum_{\mu=1}^{3} \langle V_{d_{2}}^{x_{2}\mu i_{2}} V_{\mathrm{l}}^{x_{1}\mu i_{1}}\rangle \quad d_{2} = \mathrm{l},\mathrm{c} \quad i_{2},i_{1} = 0,3,8
\label{eq_vv_corfunction}
\end{align}
and evaluate them as described in section~\ref{section2}. Operators depending on combined QCD+QED gauge links are expanded in the form $O = O{}^{(0)} + eO{}^{(\frac{1}{2})}_{e^{2}} + \frac{1}{2}e^{2}O{}^{(1)}_{e^{2}} + O(e^{3})$ making use of $W{}^{x\mu}=U{}^{x\mu}\big(\mathds{1}+\mathrm{i}eQA{}^{x\mu}-\frac{1}{2}e^{2}Q^{2}(A{}^{x\mu})^{2}\big)+O(e^{3})$. The expansion of the local vector current is trivial $(V_{\mathrm{l}})^{(0)} = V_{\mathrm{l}}$, $(V_{\mathrm{l}})^{(\frac{1}{2})}_{e^{2}}=(V_{\mathrm{l}})^{(1)}_{e^{2}}=0$ and for the conserved vector current we obtain
\begin{align*}
(V{}_{\mathrm{c}}^{x\mu i})^{(0)} &= \frac{1}{2}\big(\overline{\Psi}{}^{x+\hat{\mu}}\Lambda^{i}(\gamma^{\mu}+\mathds{1})(U^{x\mu})^{\dagger}\Psi{}^{x}+\overline{\Psi}{}^{x}\Lambda^{i}(\gamma^{\mu}-\mathds{1})U^{x\mu}\Psi{}^{x+\hat{\mu}}\big), \\
(V{}_{\mathrm{c}}^{x\mu i})^{(\frac{1}{2})}_{e^{2}} &= \frac{\mathrm{i}}{2}\big(\overline{\Psi}{}^{x+\hat{\mu}}Q\Lambda^{i}(-\gamma^{\mu}-\mathds{1})(U^{x\mu})^{\dagger}\Psi{}^{x}+\overline{\Psi}{}^{x}\Lambda^{i}Q(\gamma^{\mu}-\mathds{1})U^{x\mu}\Psi{}^{x+\hat{\mu}}\big)A^{x\mu}, \\
(V_{\mathrm{c}}^{x\mu i})^{(1)}_{e^{2}} &= -\frac{1}{2}\big(\overline{\Psi}{}^{x+\hat{\mu}}Q^{2}\Lambda^{i}(\gamma^{\mu}+\mathds{1})(U^{x\mu})^{\dagger}\Psi{}^{x}+\overline{\Psi}{}^{x}\Lambda^{i}Q^{2}(\gamma^{\mu}-\mathds{1})U^{x\mu}\Psi{}^{x+\hat{\mu}}\big)(A^{x\mu})^{2},
\end{align*}
respectively. Neglecting diagrams in which isospin-breaking is solely present in the sea quarks the diagrammatic representation of the expansion $C=C^{(0)}+\sum_{l}\Delta\varepsilon_{l}C^{(1)}_{l}+O(\Delta\varepsilon^{2})$ of Eq.~\eqref{eq_vv_corfunction} becomes
\begin{align}
\big(C{}^{i_{2}i_{1}}_{d_{2}\mathrm{l}}\big)^{(0)} &= \Big\langle
\begin{gathered}
\includegraphics[width=5.5em]{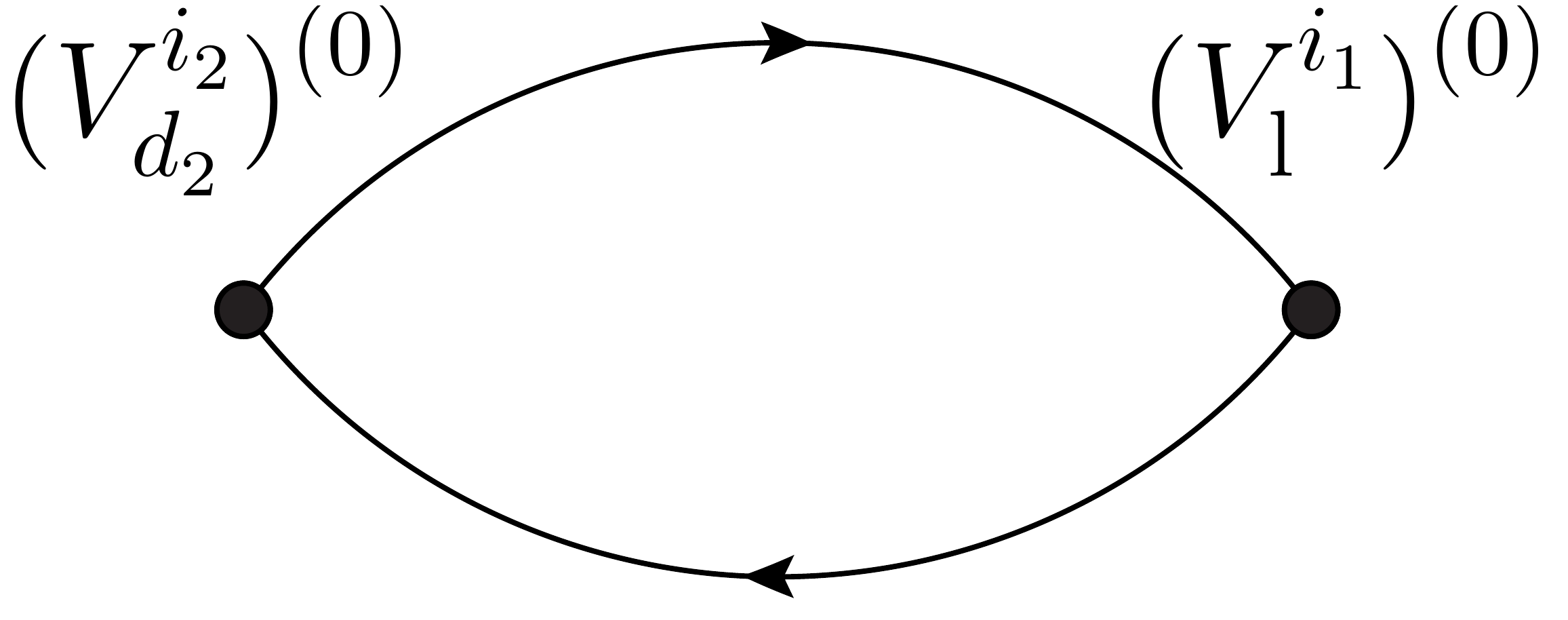}
\end{gathered} + 
\begin{gathered}
\includegraphics[width=5.5em]{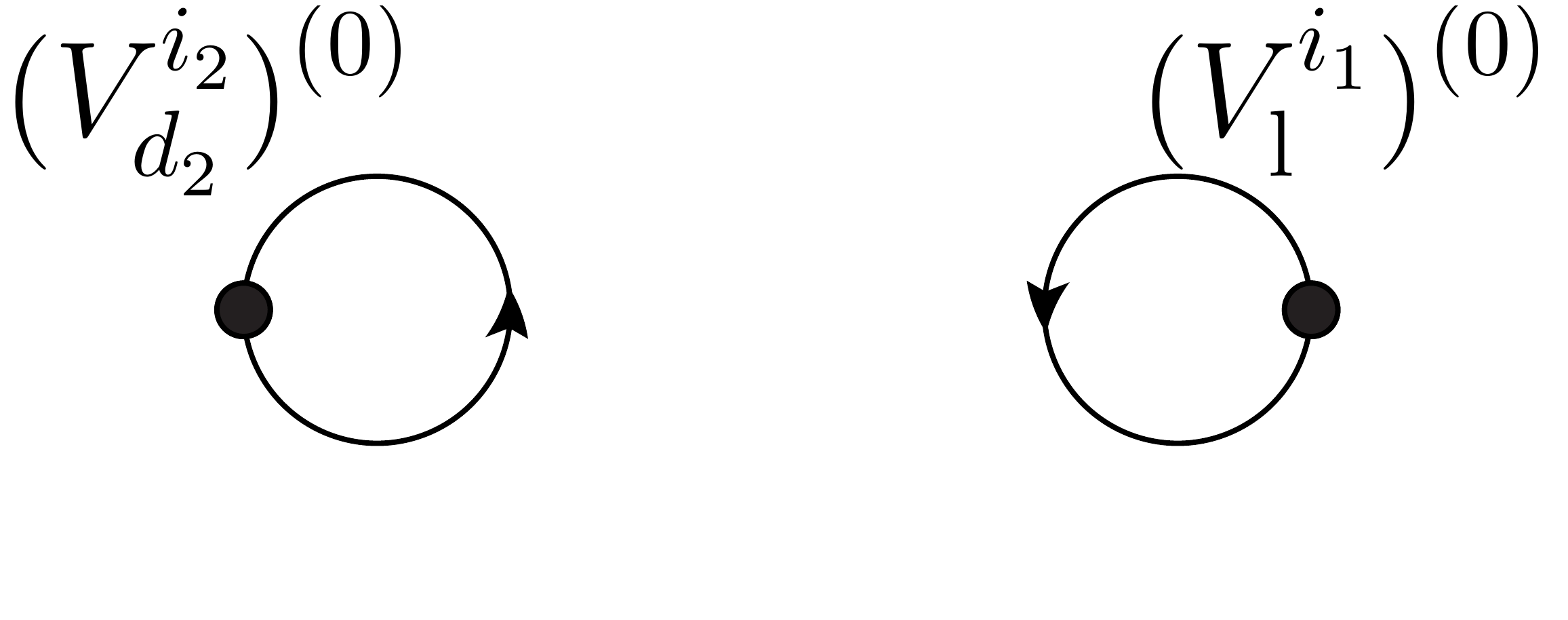}
\end{gathered}
\Big\rangle_{\mathrm{eff}}^{(0)}, \nonumber \\
\big(C{}^{i_{2}i_{1}}_{d_{2}\mathrm{l}}\big)^{(1)}_{\Delta m_{f}} &= \Big\langle
\begin{gathered}
\includegraphics[width=5.5em]{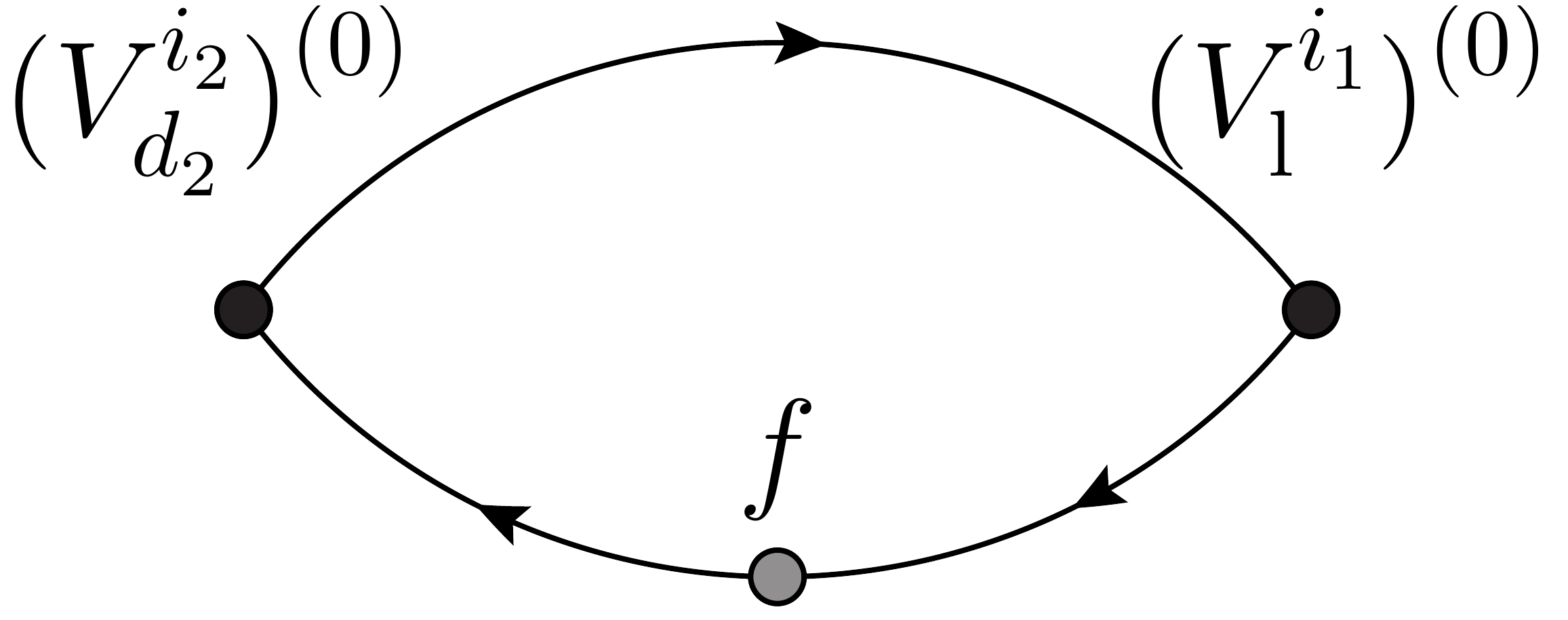}
\end{gathered}
+
\begin{gathered}
\includegraphics[width=5.5em]{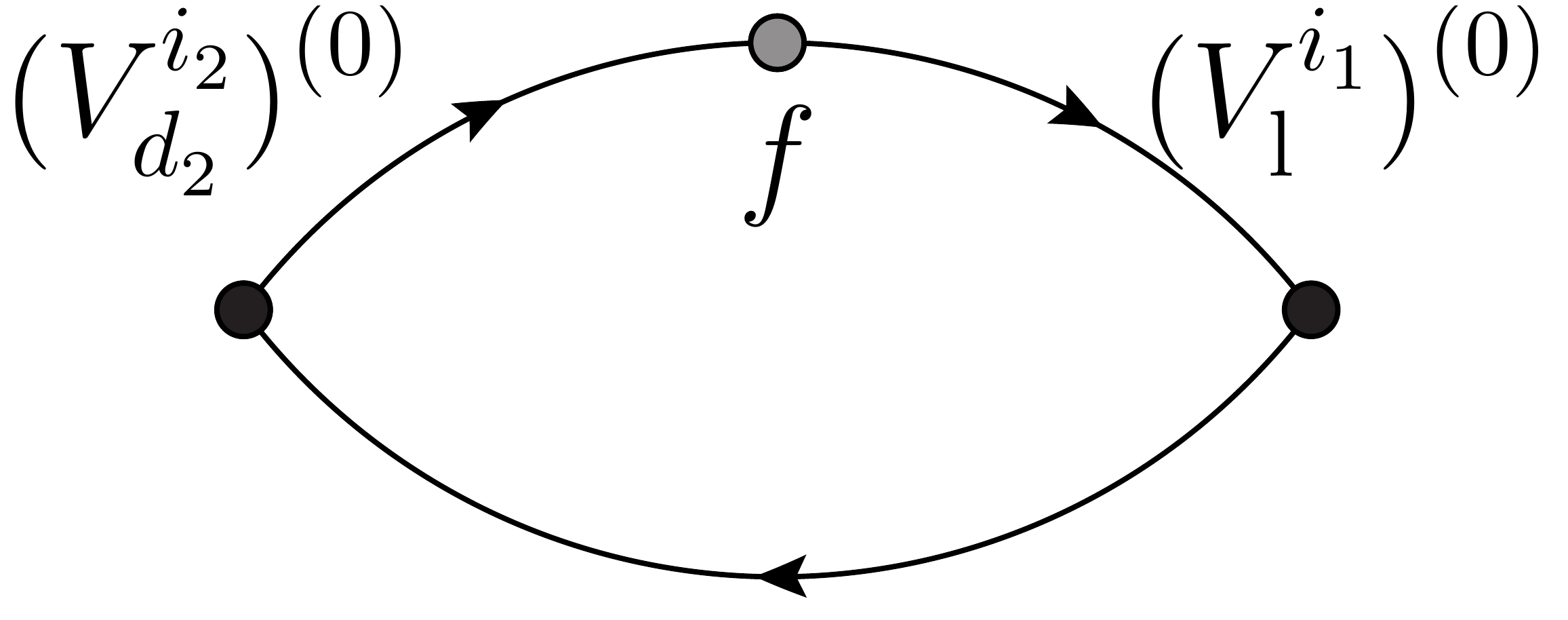}
\end{gathered}
+ 
\begin{gathered}
\includegraphics[width=5.5em]{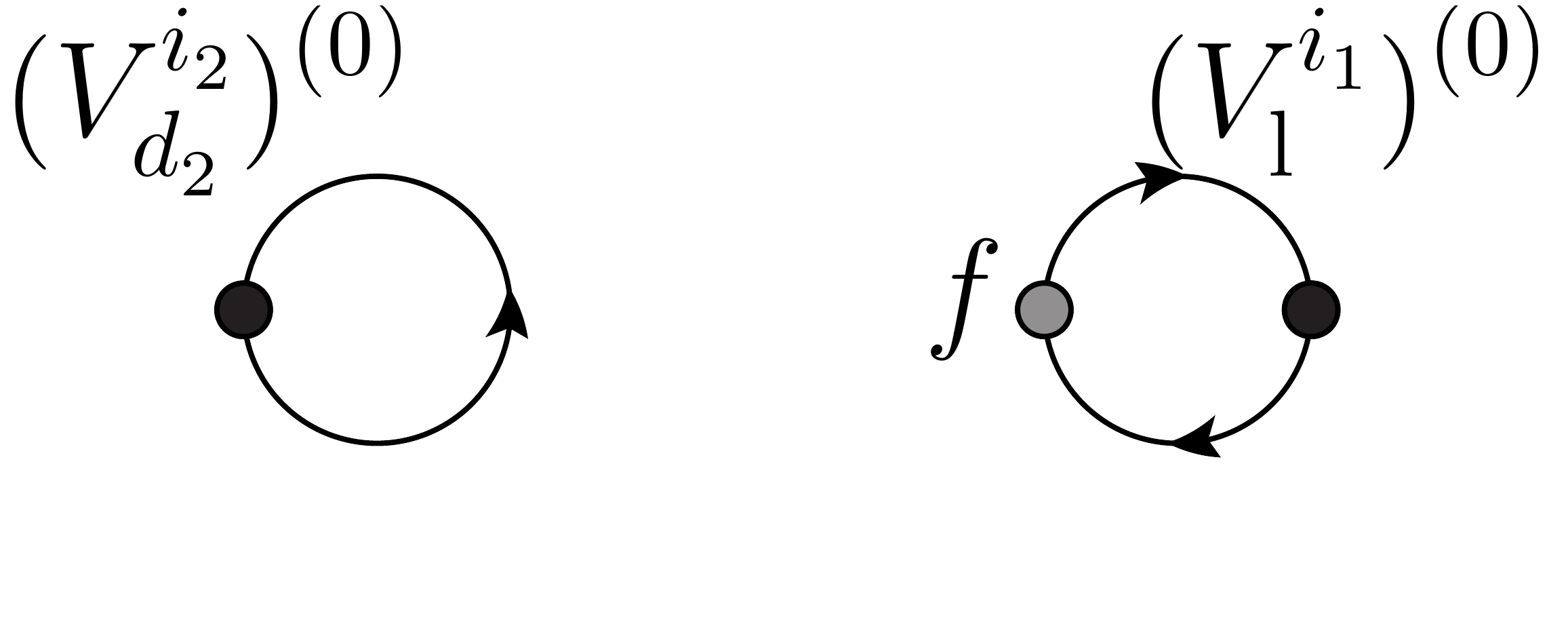}
\end{gathered}
+ 
\begin{gathered}
\includegraphics[width=5.5em]{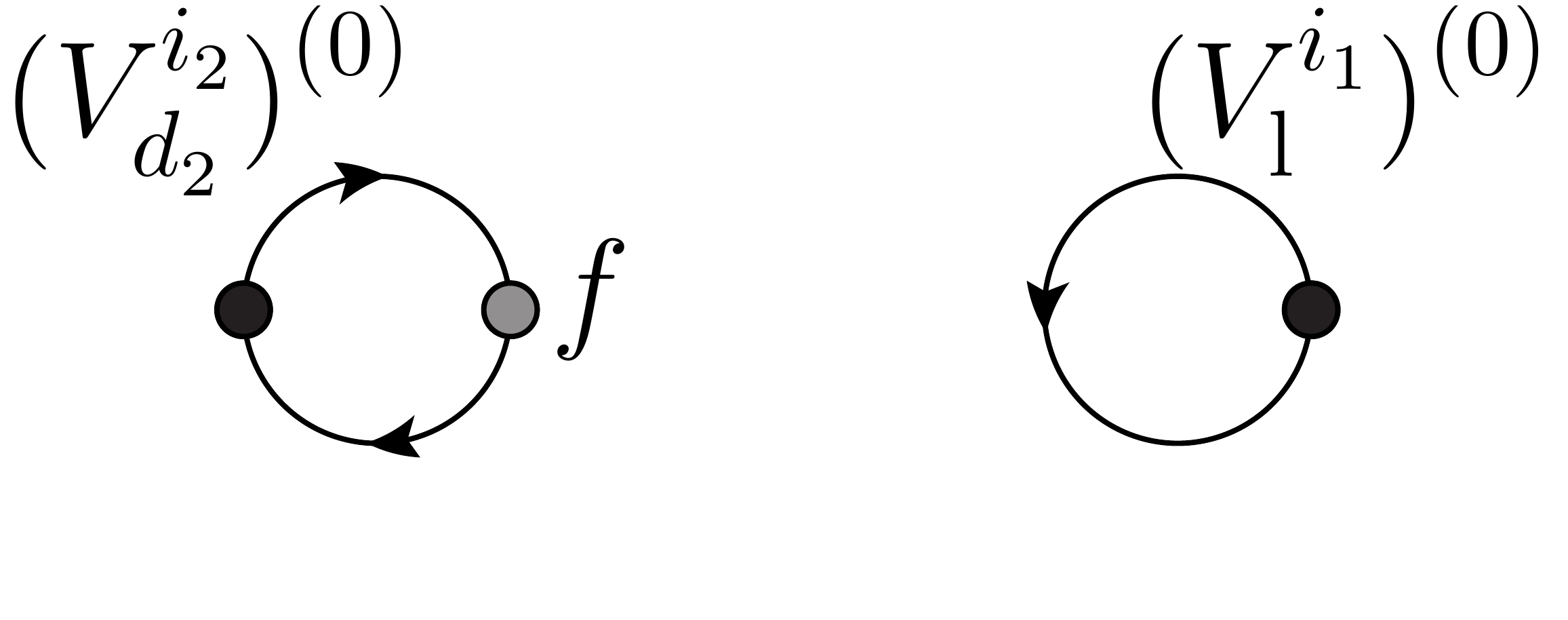}
\end{gathered}
\Big\rangle_{\mathrm{eff}}^{(0)}, \nonumber \\
\big(C{}^{i_{2}i_{1}}_{d_{2}\mathrm{l}}\big)^{(1)}_{\Delta \beta} &= \Big\langle
\begin{gathered}
\includegraphics[width=5.5em]{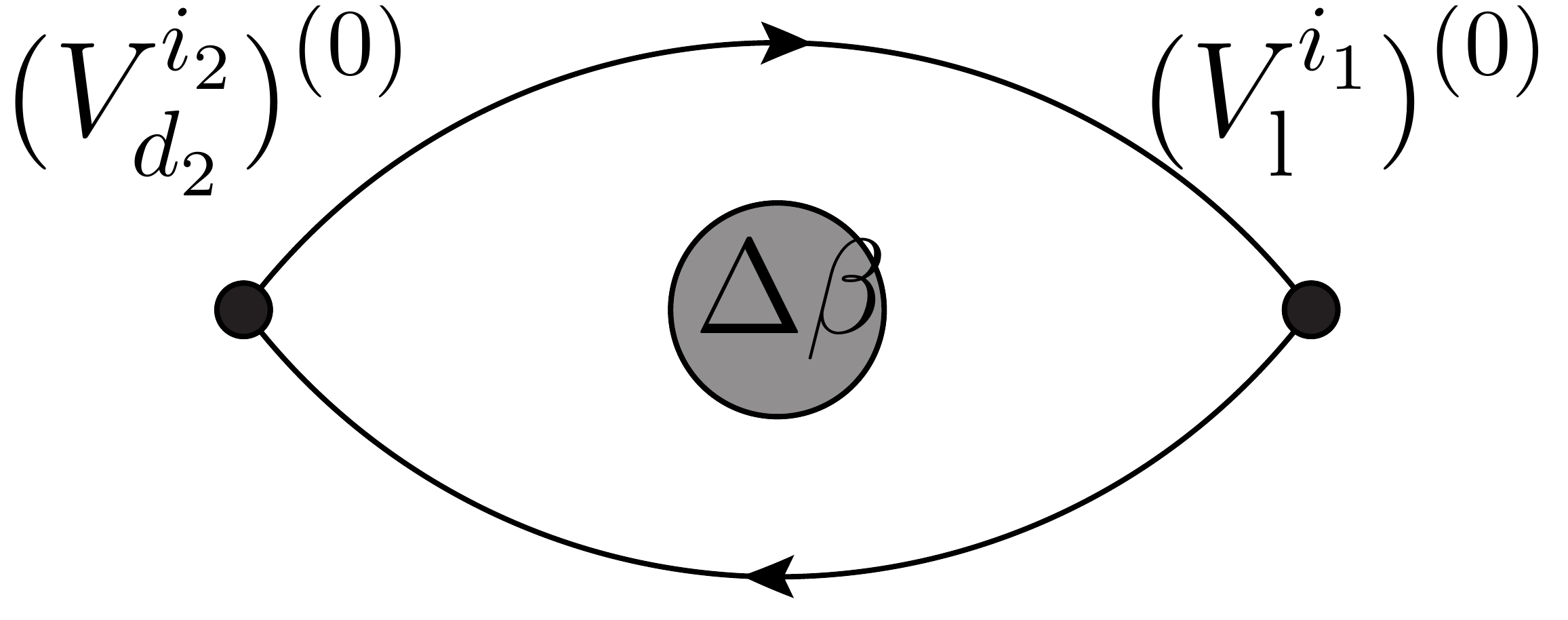}
\end{gathered} + 
\begin{gathered}
\includegraphics[width=5.5em]{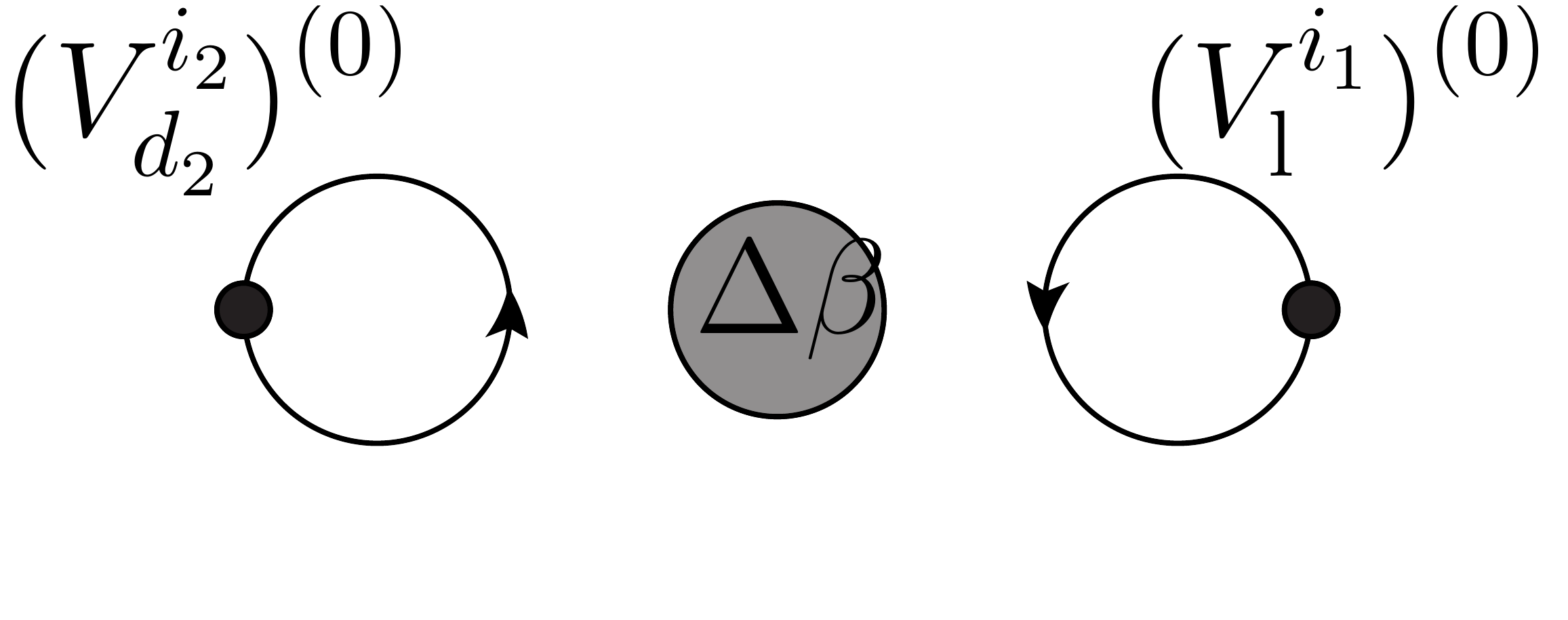}
\end{gathered}
\Big\rangle_{\mathrm{eff}}^{(0)} - \Big\langle
\begin{gathered}
\includegraphics[width=1em]{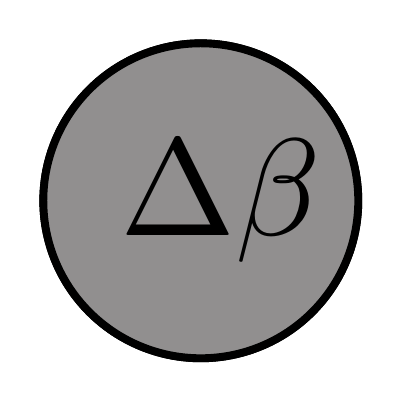}
\end{gathered}
\Big\rangle_{\mathrm{eff}}^{(0)}\Big\langle
\begin{gathered}
\includegraphics[width=5.5em]{vv_con0.pdf}
\end{gathered} + 
\begin{gathered}
\includegraphics[width=5.5em]{vv_dis0.pdf}
\end{gathered}
\Big\rangle_{\mathrm{eff}}^{(0)}, \nonumber \\
\big(C{}^{i_{2}i_{1}}_{d_{2}\mathrm{l}}\big)^{(1)}_{e^{2}} &= \Big\langle
\begin{gathered}
\includegraphics[width=5.5em]{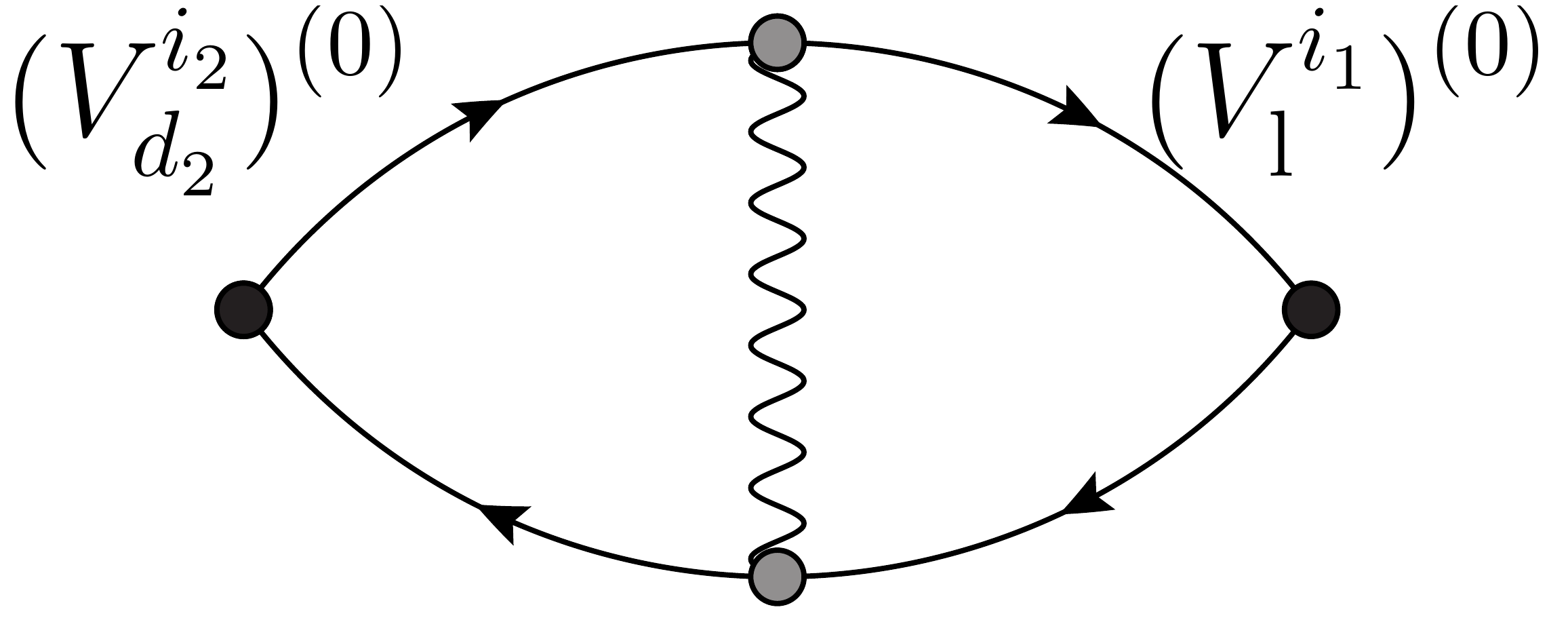}
\end{gathered}
+
\begin{gathered}
\includegraphics[width=5.5em]{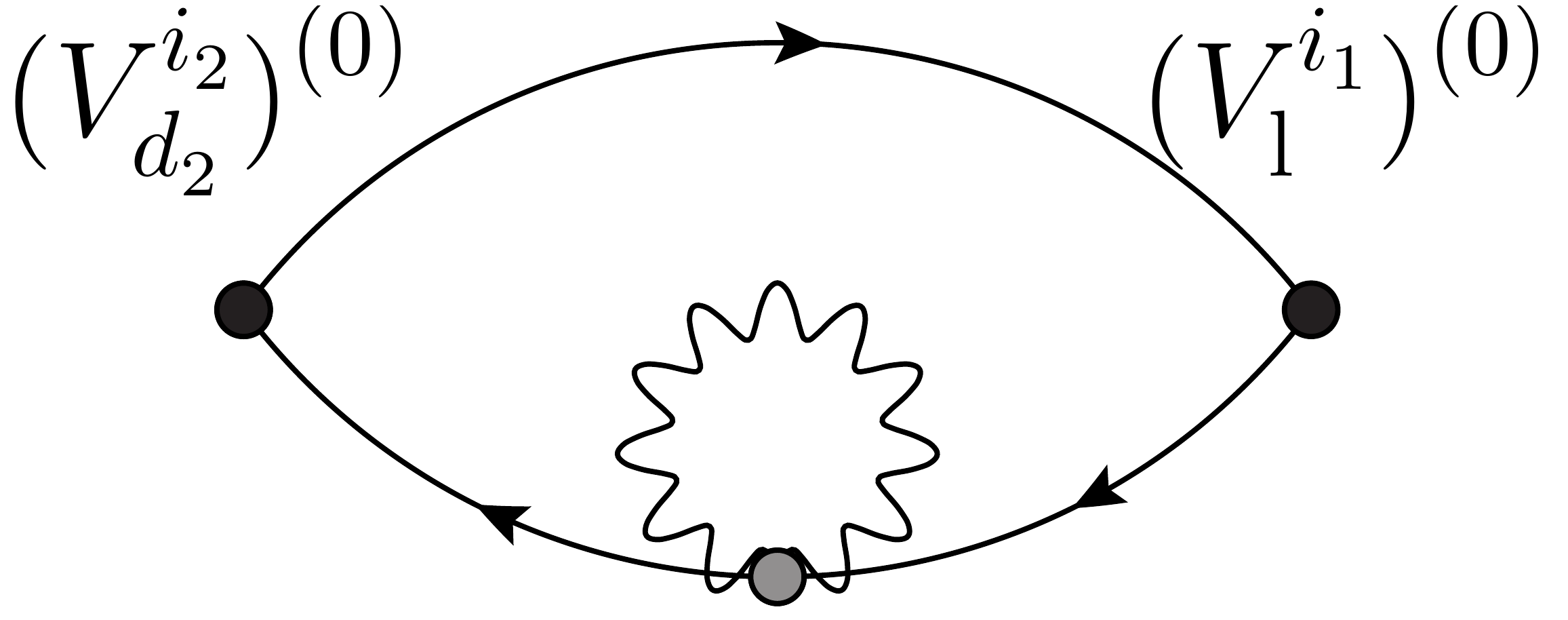}
\end{gathered}
+
\begin{gathered}
\includegraphics[width=5.5em]{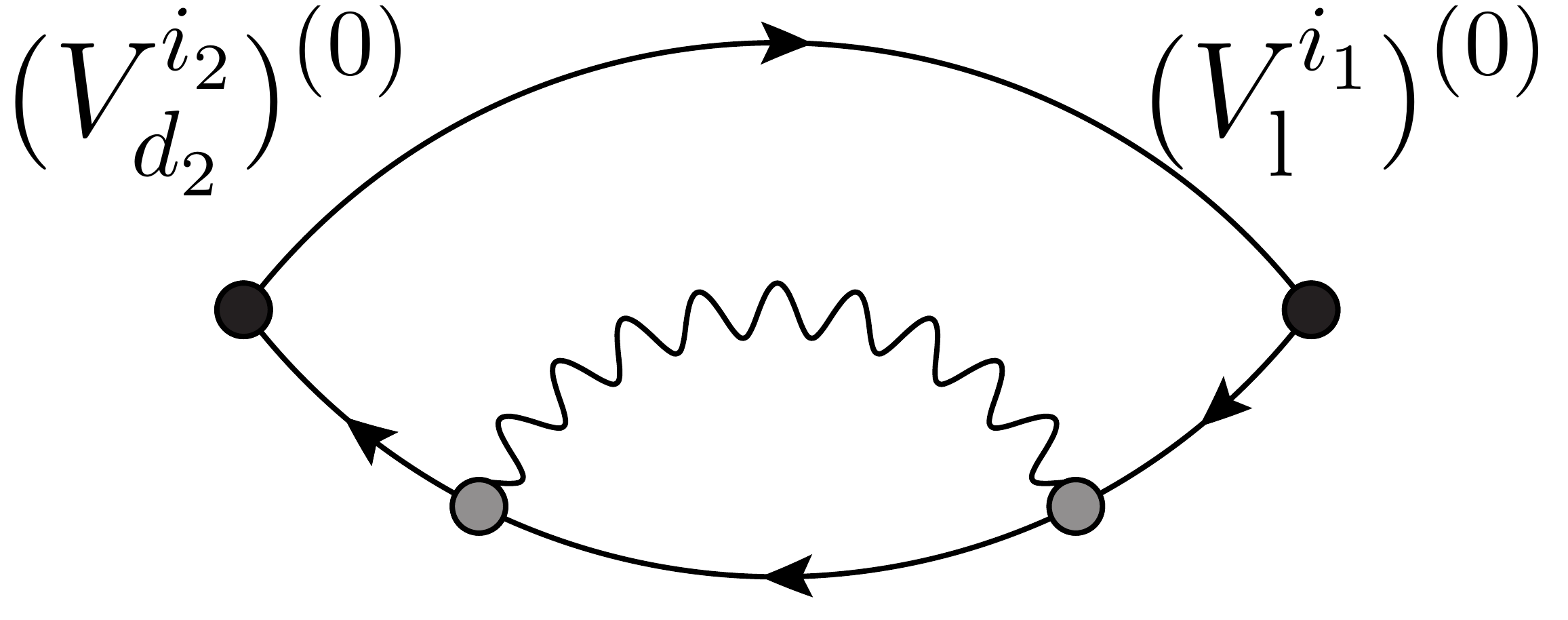}
\end{gathered}
+
\begin{gathered}
\includegraphics[width=5.5em]{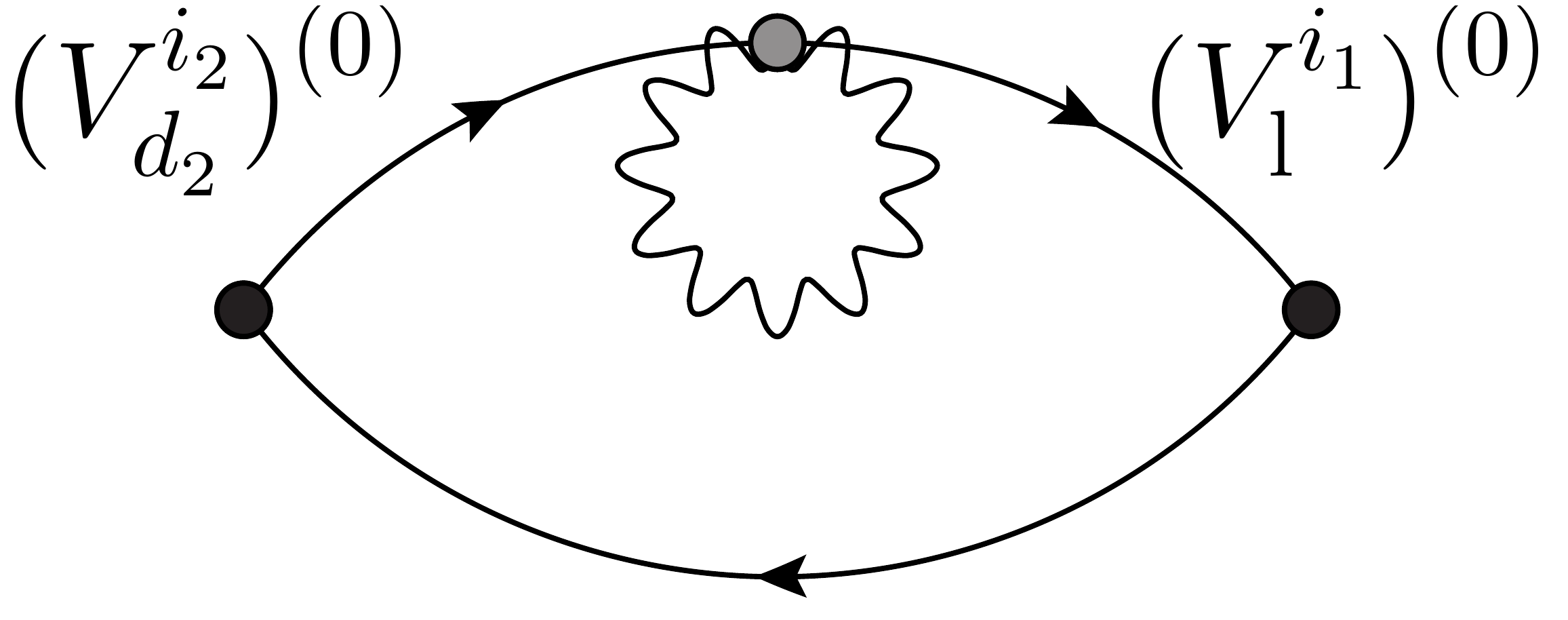}
\end{gathered}
+
\begin{gathered}
\includegraphics[width=5.5em]{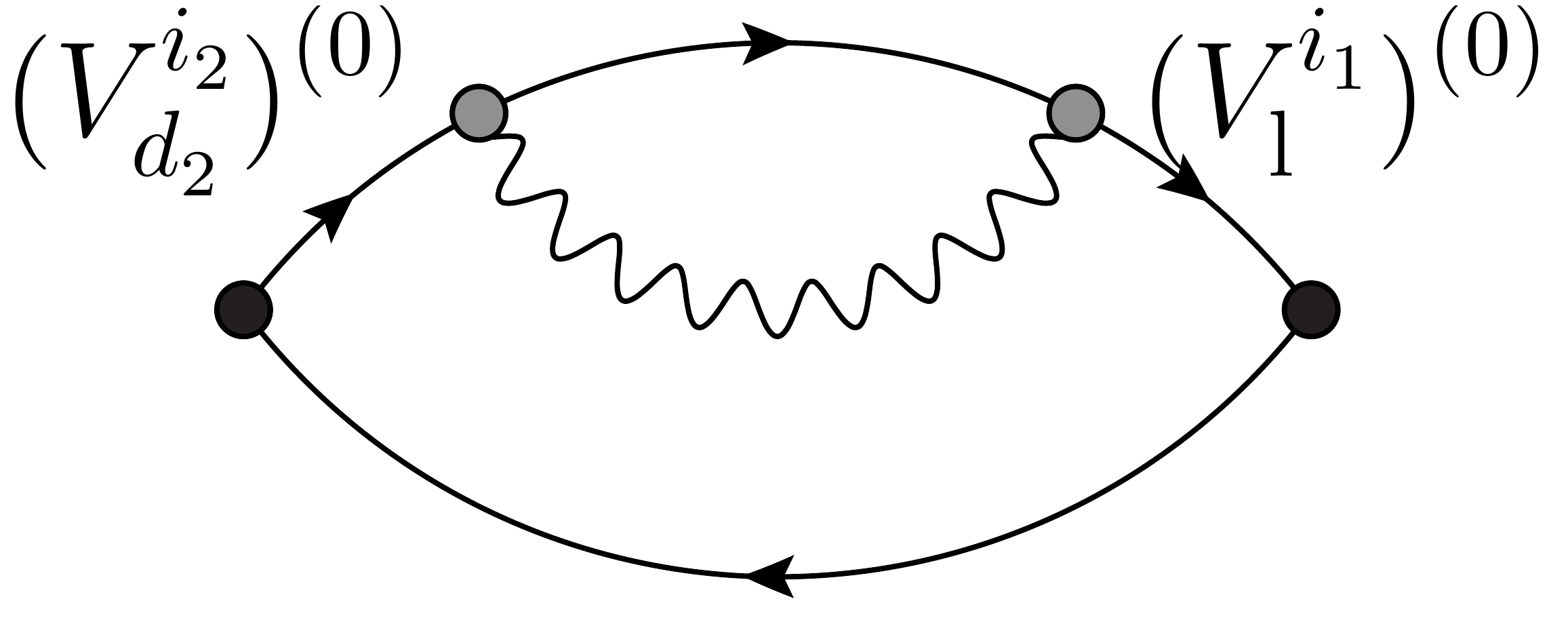}
\end{gathered} \nonumber \\
&\hphantom{\Big\langle}+
\begin{gathered}
\includegraphics[width=5.5em]{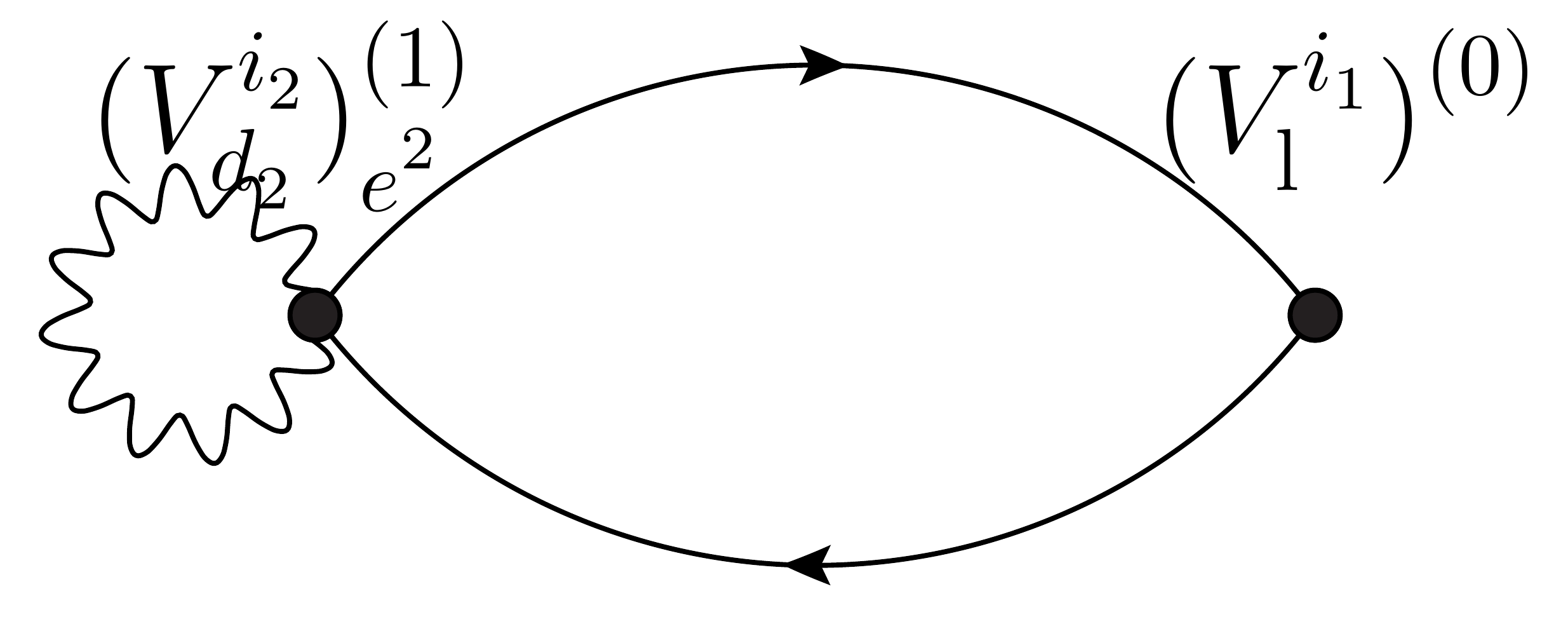}
\end{gathered}
+
\begin{gathered}
\includegraphics[width=5.5em]{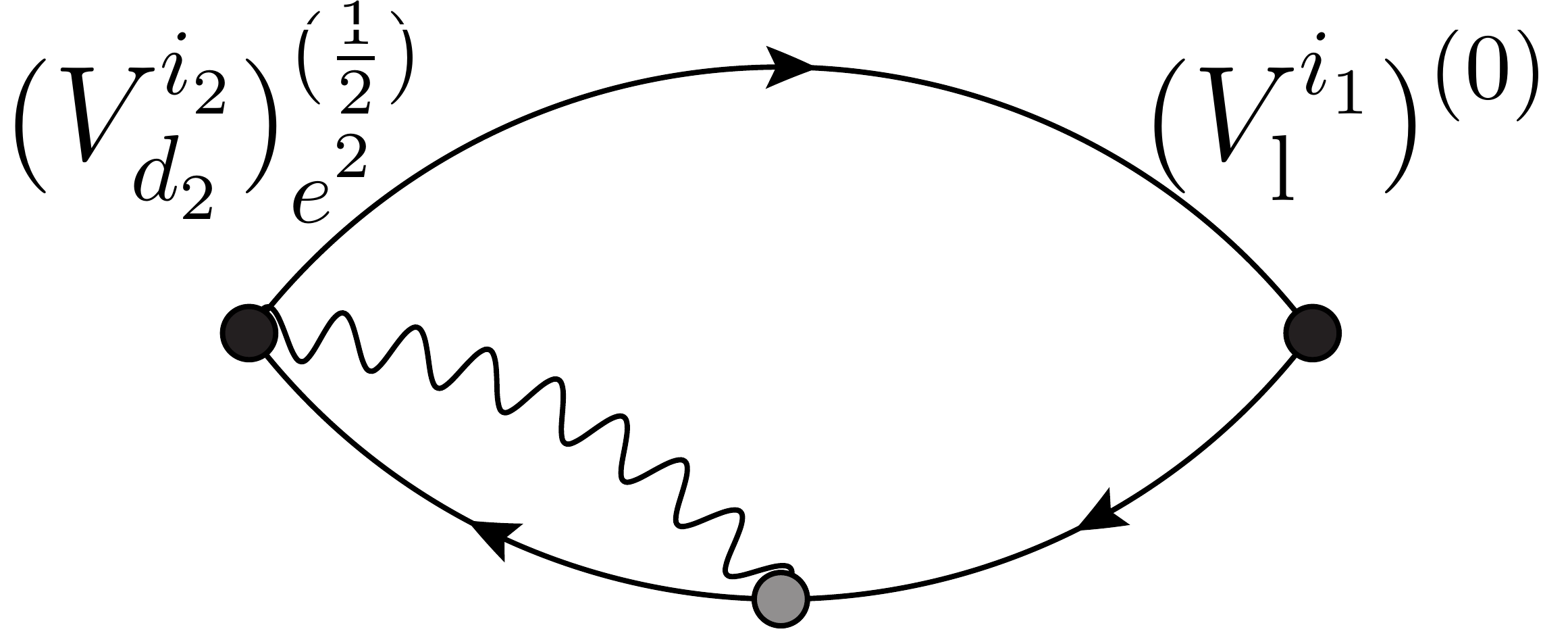}
\end{gathered}
+
\begin{gathered}
\includegraphics[width=5.5em]{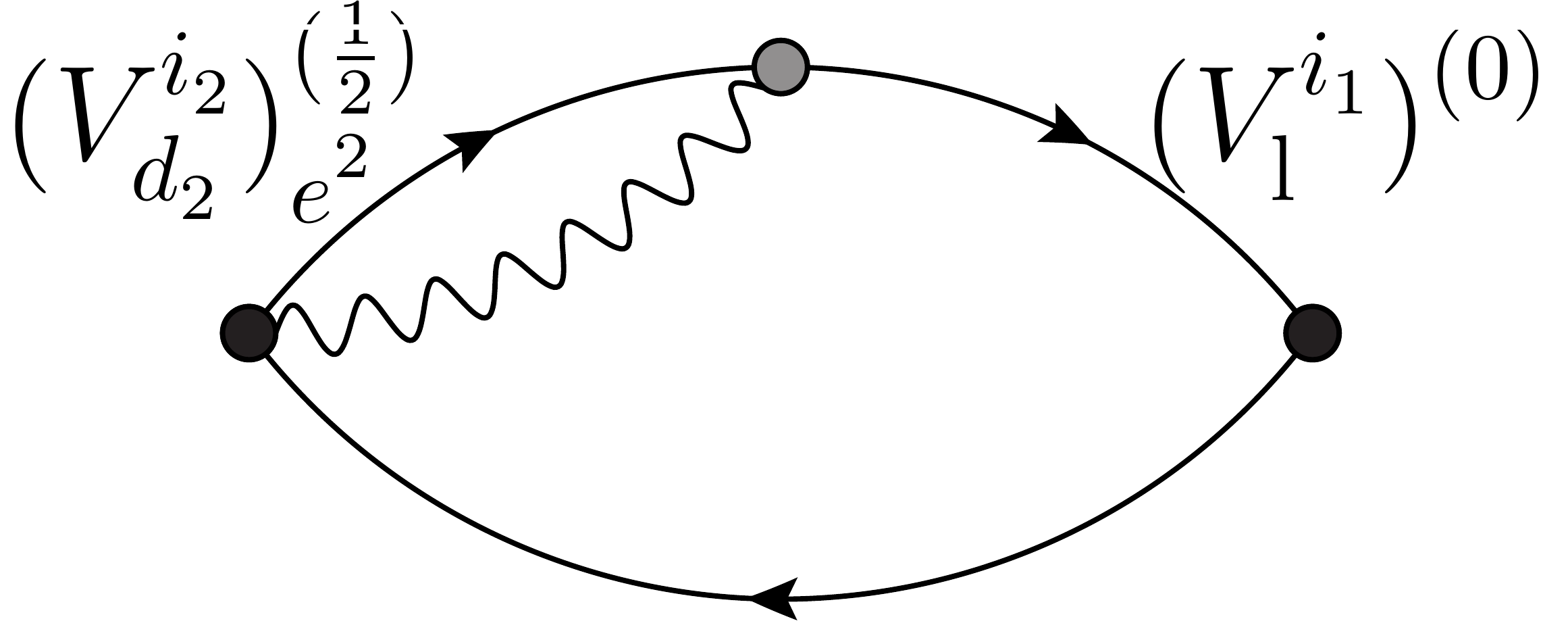}
\end{gathered}
+
\begin{gathered}
\includegraphics[width=5.5em]{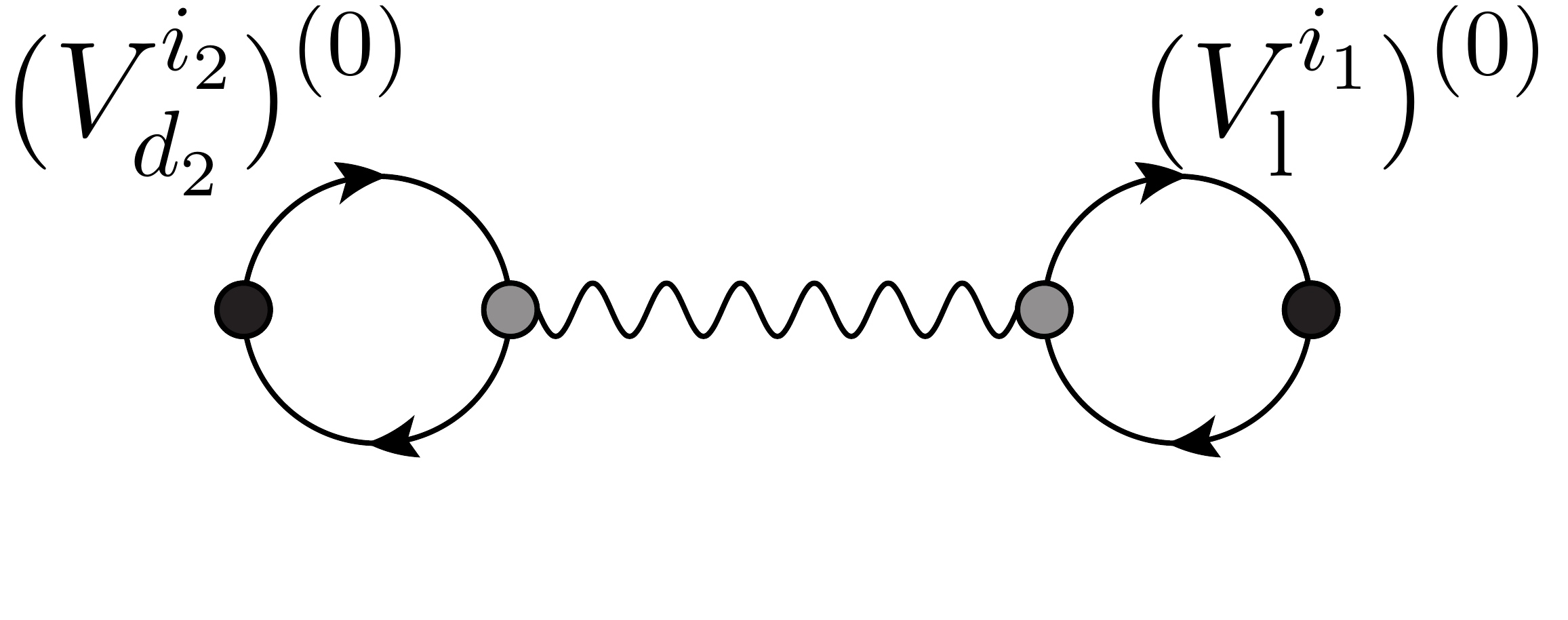}
\end{gathered}
+
\begin{gathered}
\includegraphics[width=5.5em]{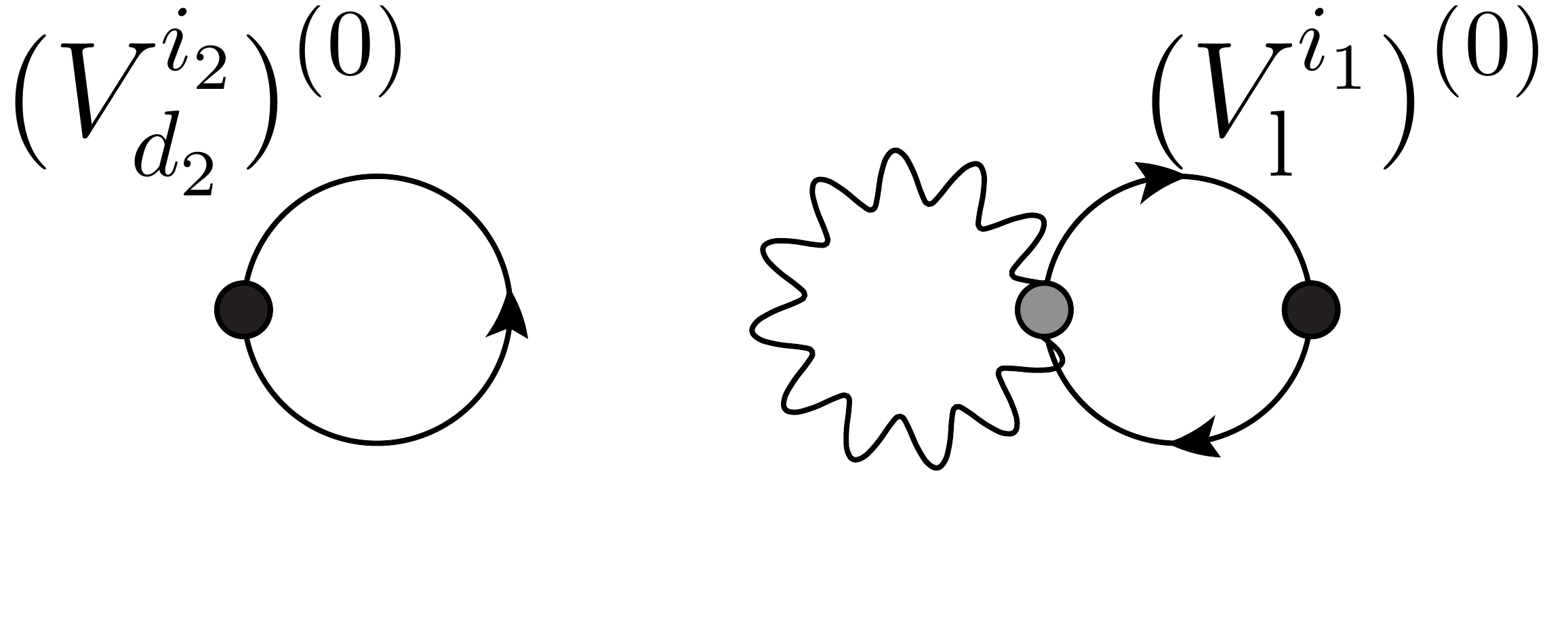}
\end{gathered} \nonumber \\
&\hphantom{\Big\langle}+
\begin{gathered}
\includegraphics[width=5.5em]{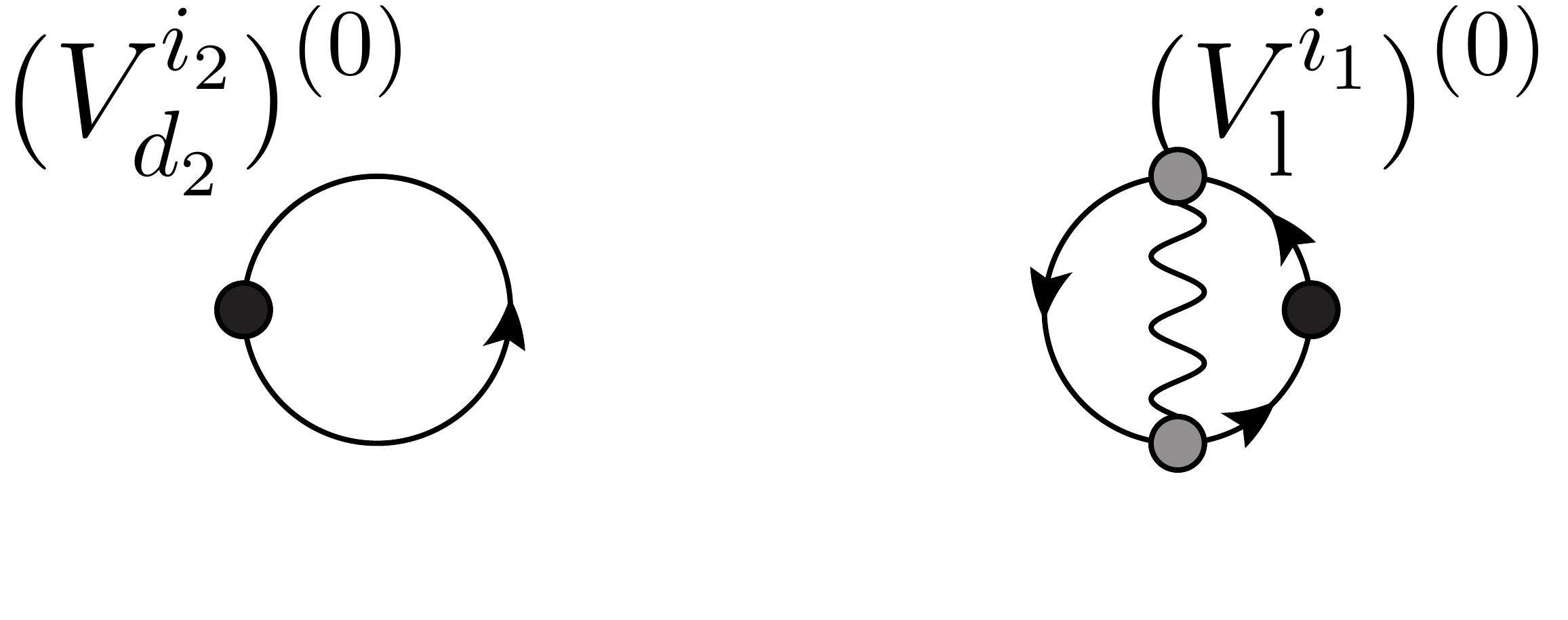}
\end{gathered}
+
\begin{gathered}
\includegraphics[width=5.5em]{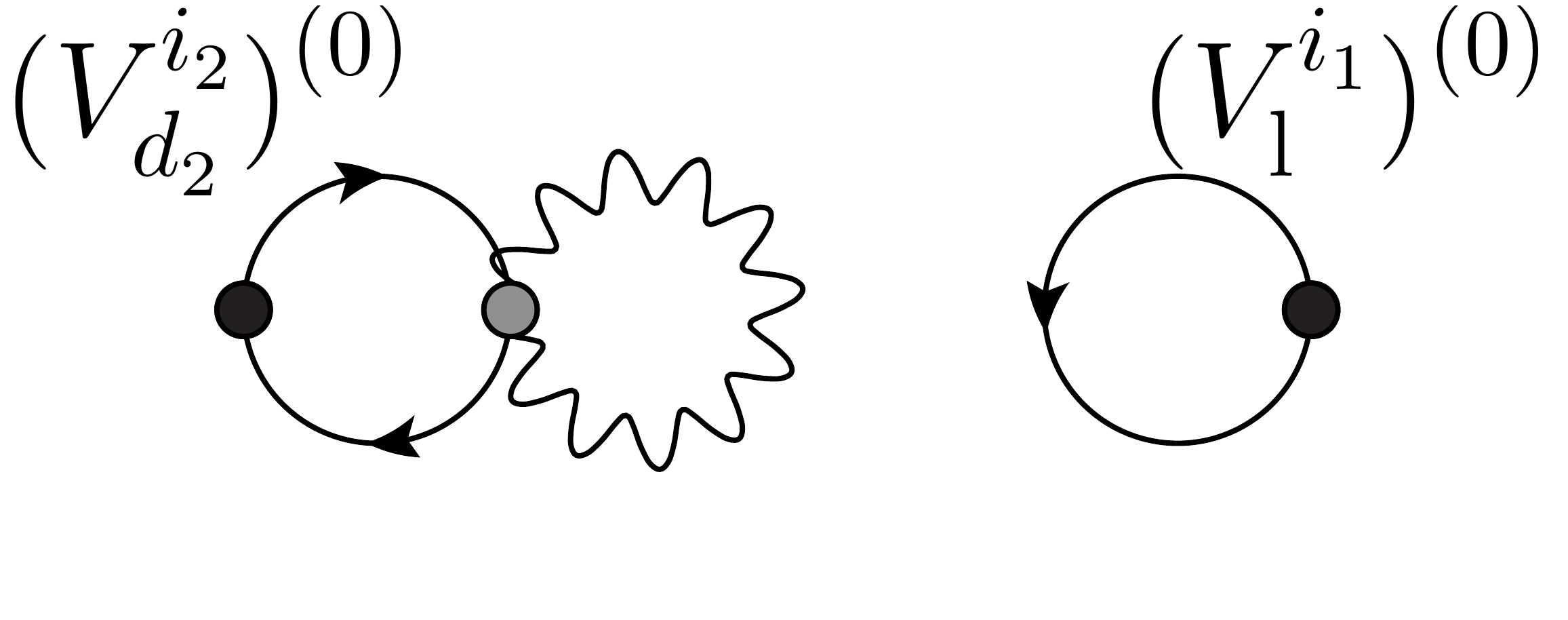}
\end{gathered}
+
\begin{gathered}
\includegraphics[width=5.5em]{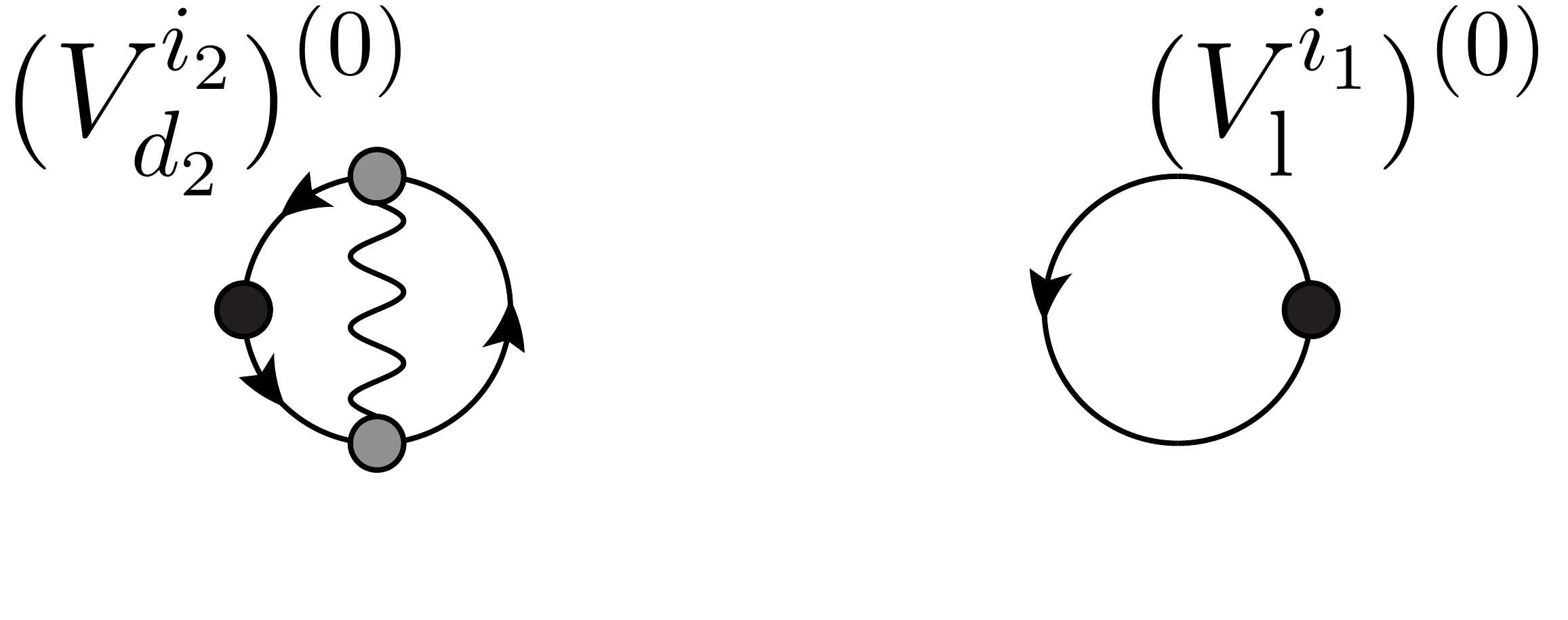}
\end{gathered}
+
\begin{gathered}
\includegraphics[width=5.5em]{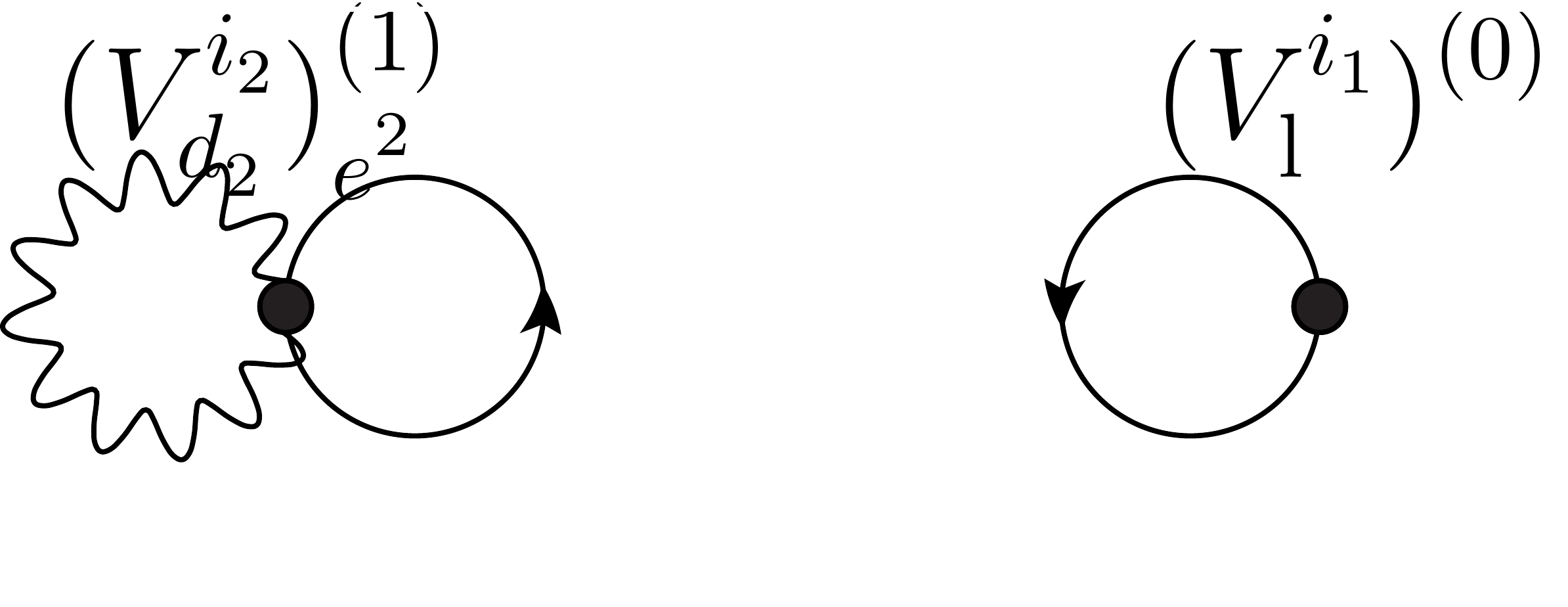}
\end{gathered}
+
\begin{gathered}
\includegraphics[width=5.5em]{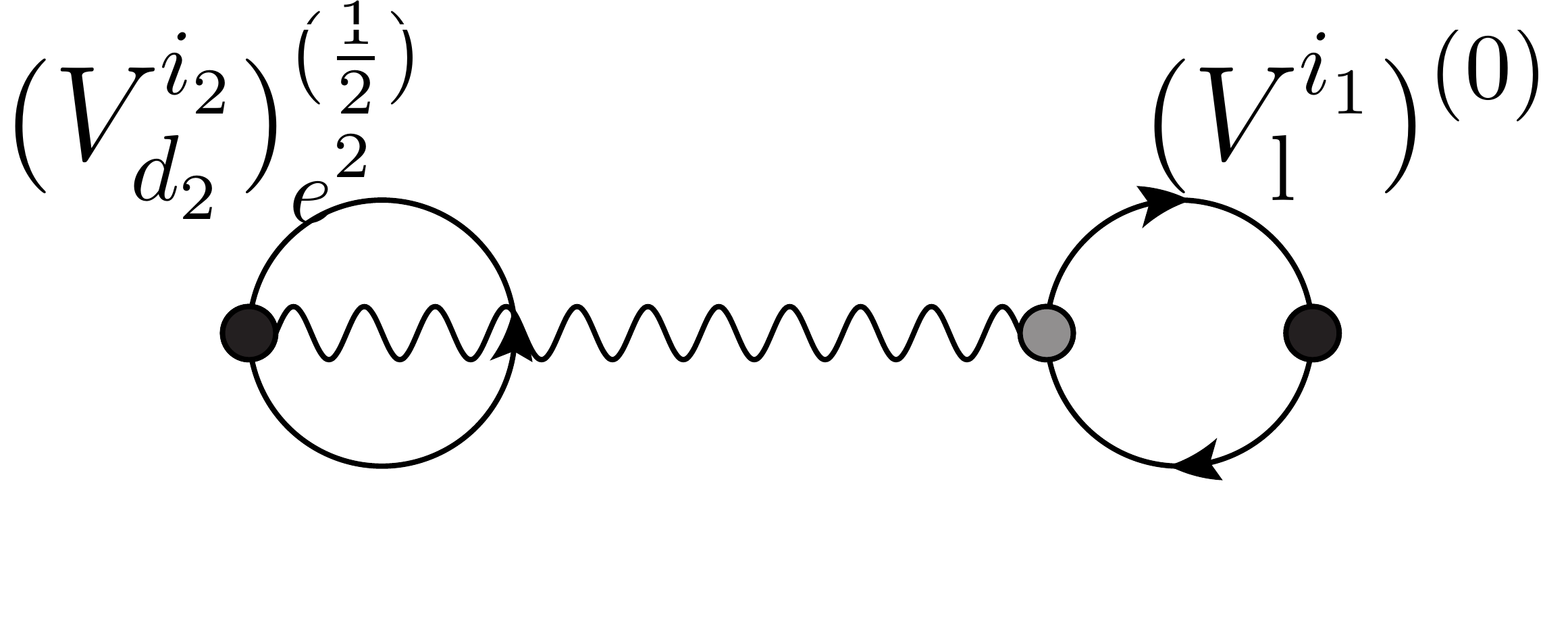}
\end{gathered} \nonumber \\
&\hphantom{\Big\langle}+
\begin{gathered}
\includegraphics[width=5.5em]{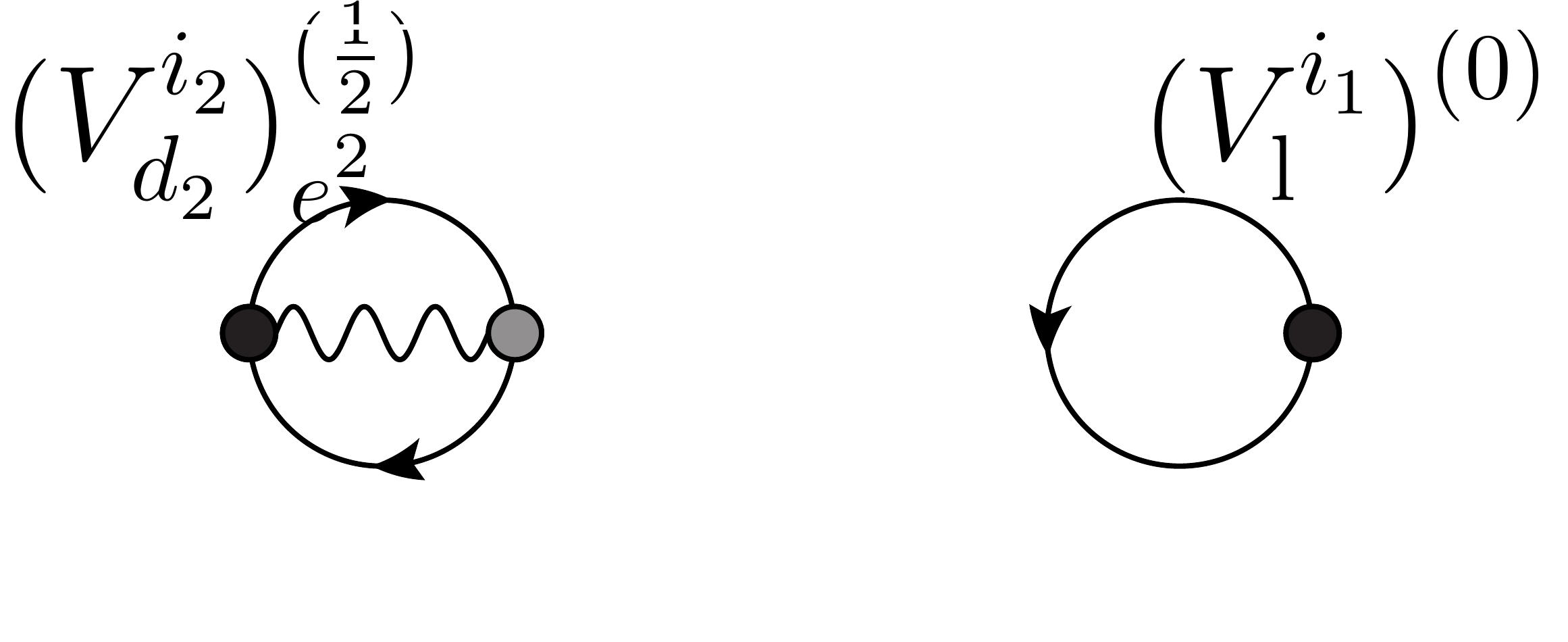}
\end{gathered}
\Big\rangle_{\mathrm{eff}}^{(0)}
\label{eq_ibvalence_diagrams}.
\end{align}

\section{Renormalisation of the local electromagnetic current}

\begin{figure}
\centering
\includegraphics[width=0.49\textwidth]{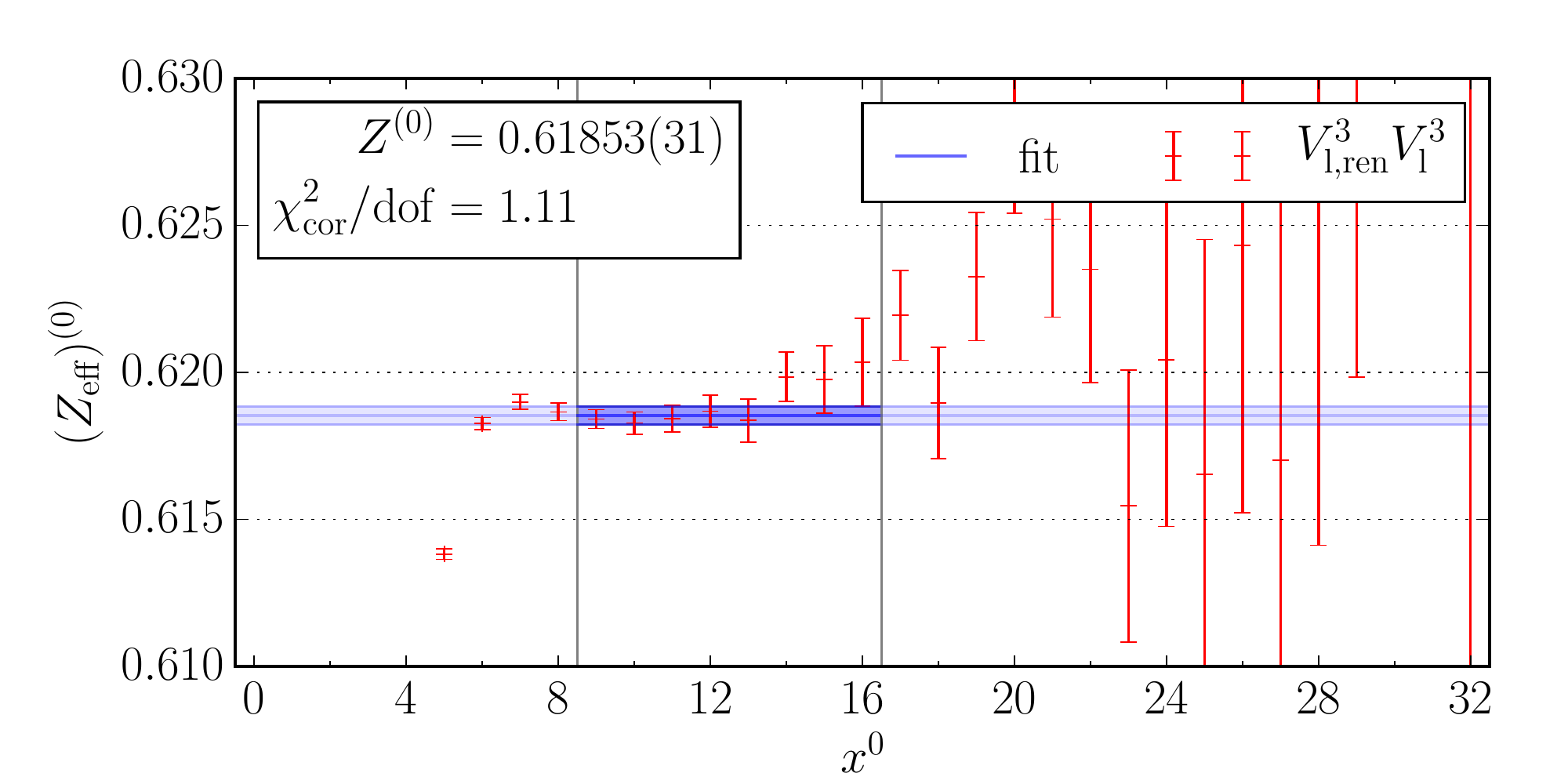}
\includegraphics[width=0.49\textwidth]{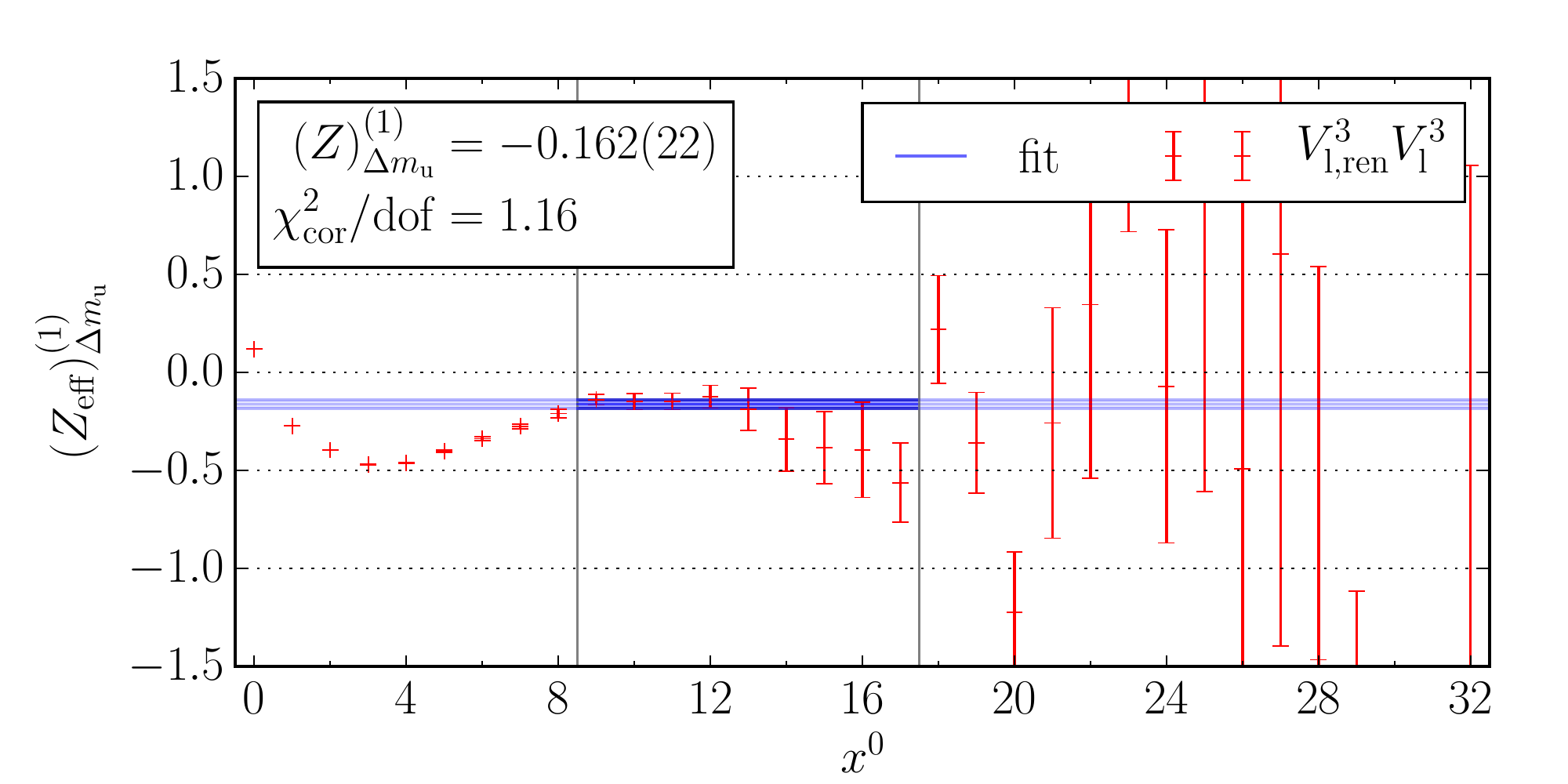}
\includegraphics[width=0.49\textwidth]{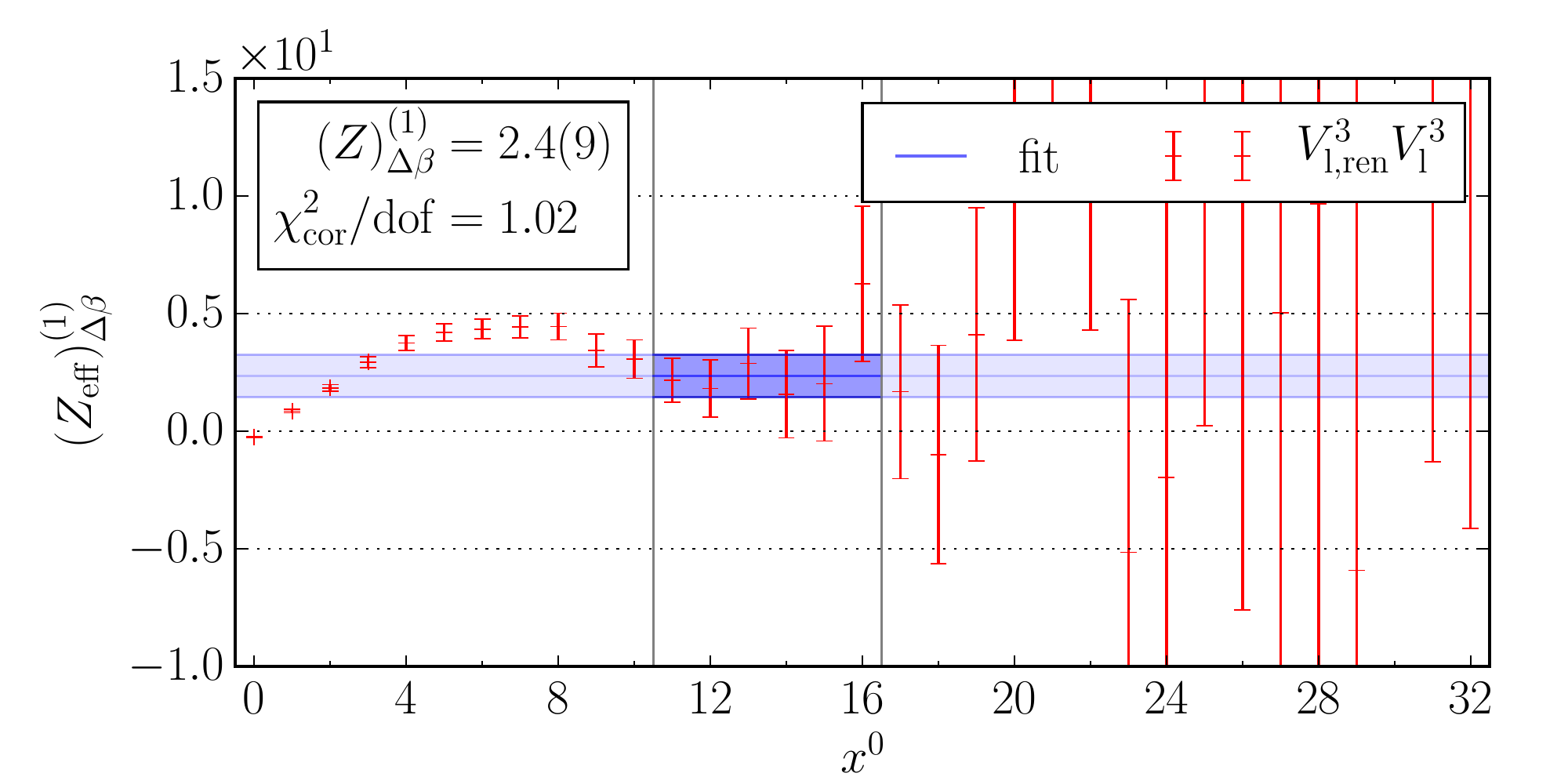}
\includegraphics[width=0.49\textwidth]{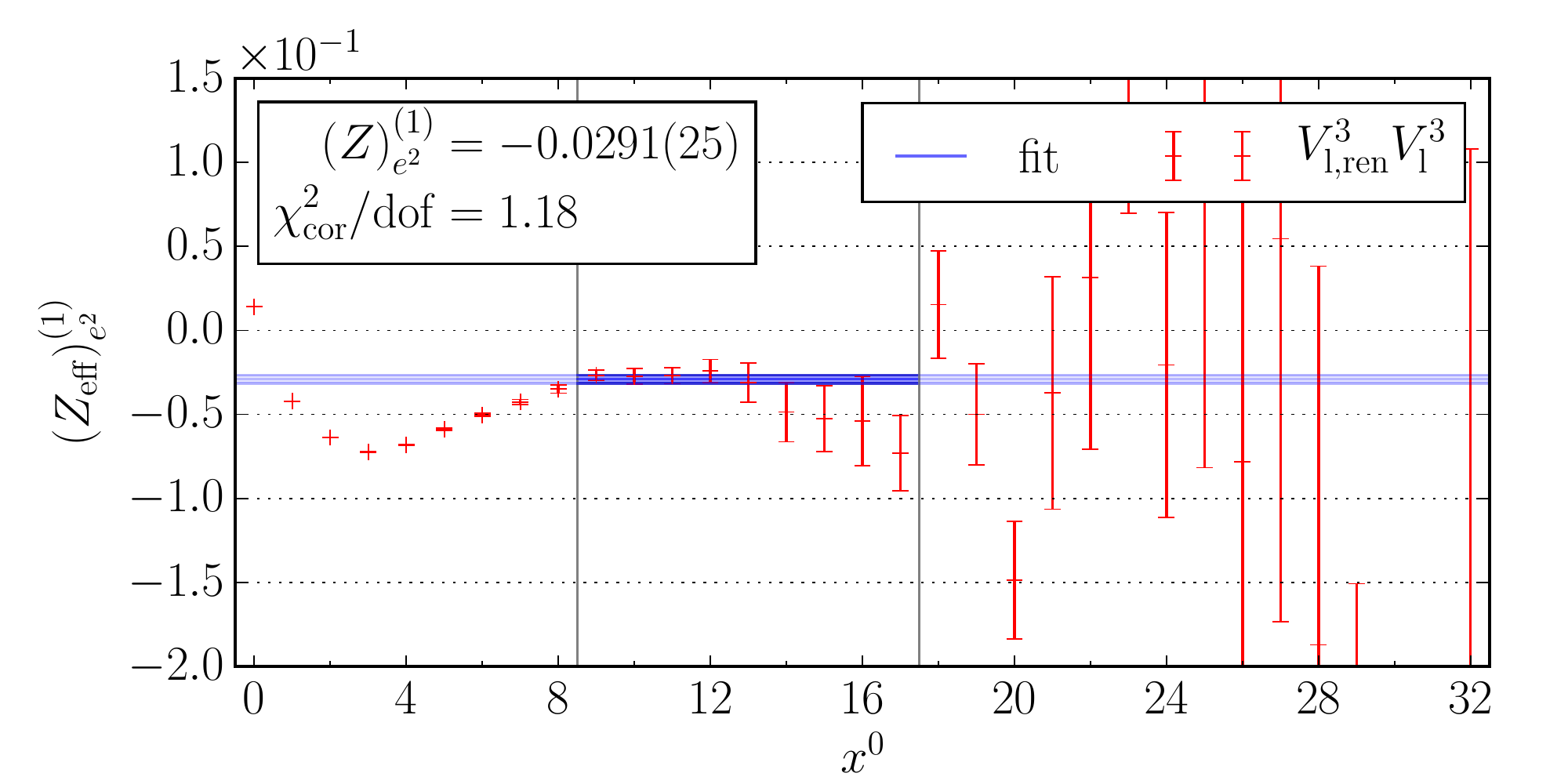}
\caption[]{Effective renormalisation factor $Z_{\mathrm{eff},V_{\mathrm{l,ren}}^{3}V_{\mathrm{l}}^{3}}$ according to Eq.~\eqref{eq_effective_renormalisation_factors} in lattice units.}
\label{fig_eff_ren_factors}
\end{figure}

In order to calculate the HVP using local electromagnetic current operators we have to determine the renormalisation pattern of $V_{\mathrm{l}}^{\gamma}$. Treating isospin-breaking effects perturbatively it is reasonable to consider an operator basis with definite transformation behaviour under isospin rotations. For a given $\mu$ the flavour neutral vector currents $V_{\mathrm{l}}^{i\mu}$ with $i=0,3,8$ exhibit mixing under renormalisation, such that we have to introduce a matrix of renormalisation factors with entries $Z_{V_{\mathrm{l,ren}}^{i_{2}}V_{\mathrm{l}}^{i_{1}}}$ and $i_{2},i_{1}=0,3,8$ to ensure the correct multiplicative renormalisation $V_{\mathrm{l,ren}} = Z_{V_{\mathrm{l,ren}}V_{\mathrm{l}}} V_{\mathrm{l}}$. Although open boundary conditions break the translational invariance in time direction, we expect the renormalisation factor to have negligible time dependency far away from the boundaries. Correctly renormalised correlation functions converge to their continuum limit with a rate defined by a power in the lattice spacing $a$. Considering that the conserved vector current $V_{\mathrm{c}}$ does not renormalise, i.e. $Z_{V_{\mathrm{c,ren}}V_{\mathrm{c}}}=\mathds{1}$, we define a renormalisation condition for the local vector current $V_{\mathrm{l}}$:
\begin{align}
\langle V_{\mathrm{c},\mathrm{ren}} V_{\mathrm{l},\mathrm{ren}}\rangle - \langle V_{\mathrm{l},\mathrm{ren}} V_{\mathrm{l},\mathrm{ren}}\rangle =  Z_{V_{\mathrm{c,ren}}V_{\mathrm{c}}}\langle V_{\mathrm{c}} V_{\mathrm{l}}\rangle Z_{V_{\mathrm{l,ren}}V_{\mathrm{l}}}{}^{T} - Z_{V_{\mathrm{l,ren}}V_{\mathrm{l}}} \langle V_{\mathrm{l}} V_{\mathrm{l}}\rangle Z_{V_{\mathrm{l,ren}}V_{\mathrm{l}}}{}^{T} &= O(a) \nonumber\\
\Leftrightarrow \langle V_{\mathrm{c}} V_{\mathrm{l}}\rangle - Z_{V_{\mathrm{l,ren}}V_{\mathrm{l}}} \langle V_{\mathrm{l}} V_{\mathrm{l}}\rangle &= O(a).\label{eq_ren_condition}
\end{align}
The $O(a)$-term is potentially time-dependent but we expect it to become constant for larger time separations~\cite{Boyle:2017gzv} as the specific form of the discretisation of the operators becomes less relevant. After averaging Eq.~\eqref{eq_ren_condition} over the spatial components and projecting to $\vec{0}$-momentum we can express the latter in terms of the correlation functions defined in Eq.~\eqref{eq_vv_corfunction}, reading $C_{\mathrm{c}\mathrm{l}} - Z_{V_{\mathrm{l,ren}}V_{\mathrm{l}}}C_{\mathrm{l}\mathrm{l}} = O(a)$. In order to extract the renormalisation constants from the correlation functions $C_{\mathrm{l}\mathrm{l}}$ and $C_{\mathrm{c}\mathrm{l}}$ we define effective time dependent renormalisation factors $Z_{\mathrm{eff},V_{\mathrm{l,ren}}V_{\mathrm{l}}} = C_{\mathrm{c}\mathrm{l}}\big(C_{\mathrm{l}\mathrm{l}}\big)^{-1}$. We perturbatively expand this definition and extract the effective renormalisation factors order by order:
\begin{align}
\big(Z_{\mathrm{eff},V_{\mathrm{l,ren}}V_{\mathrm{l}}}\big)^{(0)} &= (C_{\mathrm{c}\mathrm{l}})^{(0)}\big((C_{\mathrm{l}\mathrm{l}})^{(0)}\big)^{-1}, \nonumber \\
(Z_{\mathrm{eff},V_{\mathrm{l,ren}}V_{\mathrm{l}}})^{(1)}_{l} &= \Big((C_{\mathrm{c}\mathrm{l}})^{(1)}_{l} - (C_{\mathrm{c}\mathrm{l}})^{(0)}\big((C_{\mathrm{l}\mathrm{l}})^{(0)}\big)^{-1} (C_{\mathrm{l}\mathrm{l}})^{(1)}_{l}\Big)\big((C_{\mathrm{l}\mathrm{l}})^{(0)}\big)^{-1}. \label{eq_effective_renormalisation_factors}
\end{align}
In Fig.~\ref{fig_eff_ren_factors} results for the effective renormalisation factor $Z_{V_{\mathrm{l,ren}}^{3}V_{\mathrm{l}}^{3},\mathrm{eff}}$ are displayed, where we have only taken into account the quark-connected diagrams in Eq.~\eqref{eq_ibvalence_diagrams}. From the fits we extract the following renormalisation factors with $Z_{V_{\mathrm{l,ren}}V_{\mathrm{l}}}=(Z_{V_{\mathrm{l,ren}}V_{\mathrm{l}}})^{(0)}+\sum_{l}\Delta\varepsilon_{l}(Z_{V_{\mathrm{l,ren}}V_{\mathrm{l}}})^{(1)}_{l}+O(\Delta\varepsilon^{2})$:
\begin{align*}
(Z_{V_{\mathrm{l,ren}}V_{\mathrm{l}}})^{(0)} &=
\begin{tiny}
\begin{pmatrix}
0.61736(21) & 0.0 & 0.00159(12) \\
0.0 & 0.61853(31) & 0.0 \\
0.00159(12) & 0.0 & 0.61587(22)
\end{pmatrix}
\end{tiny},
&
(Z_{V_{\mathrm{l,ren}}V_{\mathrm{l}}})^{(1)}_{\Delta m_{\mathrm{u}}} &=
\begin{tiny}
\begin{pmatrix}
-0.092(21) & -0.113(25) & -0.065(15) \\
-0.113(25) & -0.162(22) & -0.080(18) \\
-0.065(15) & -0.080(18) & -0.046(10)
\end{pmatrix}
\end{tiny}, \\
(Z_{V_{\mathrm{l,ren}}V_{\mathrm{l}}})^{(1)}_{\Delta m_{\mathrm{d}}} &= 
\begin{tiny}
\begin{pmatrix}
-0.092(21) & 0.113(25) & -0.065(15) \\
0.113(25) & -0.162(22) & 0.080(18) \\
-0.065(15) & 0.080(18) & -0.046(10)
\end{pmatrix}
\end{tiny},
&
(Z_{V_{\mathrm{l,ren}}V_{\mathrm{l}}})^{(1)}_{\Delta m_{\mathrm{s}}} &= 
\begin{tiny}
\begin{pmatrix}
-0.149(5) & 0.0 & 0.210(7) \\
0.0 & 0.0 & 0.0 \\
0.210(7) & 0.0 & -0.298(10)
\end{pmatrix}
\end{tiny}, \\
(Z_{V_{\mathrm{l,ren}}V_{\mathrm{l}}})^{(1)}_{\Delta\beta} &= 
\begin{tiny}
\begin{pmatrix}
2.6(7) & 0.0 & -0.34(26) \\
0.0 & 2.4(9) & 0.0 \\
-0.34(26) & 0.0 & 2.7(7)
\end{pmatrix}
\end{tiny},
&
(Z_{V_{\mathrm{l,ren}}V_{\mathrm{l}}})^{(1)}_{e^{2}} &= 
\begin{tiny}
\begin{pmatrix}
-0.0225(24) & -0.0129(17) & -0.0054(16) \\
-0.0129(17) & -0.0291(25) & -0.0091(12) \\
-0.0054(16) & -0.0091(12) & -0.0186(13)
\end{pmatrix}
\end{tiny}.
\end{align*}
Vanishing components of $(Z_{V_{\mathrm{l,ren}}V_{\mathrm{l}}})^{(1)}_{\Delta m_{\mathrm{s}}}$ only receive contributions from quark-disconnected diagrams which have been neglected. The renormalised electromagnetic current in terms of the bare vector currents defined in Eq.~\eqref{eq_v_currents} is then finally given by
\begin{align}
V{}_{d,\mathrm{ren}}^{\gamma} &= V{}_{d,\mathrm{ren}}^{3} + \frac{1}{\sqrt{3}} V{}_{d,\mathrm{ren}}^{8} = \sum_{i} \Big(Z_{V_{d,\mathrm{ren}}^{3}V_{d}^{i}} + \frac{1}{\sqrt{3}} Z_{V_{d,\mathrm{ren}}^{8}V_{d}^{i}}\Big) V{}_{d}^{i} & d &= \mathrm{l},\mathrm{c}.
\label{eq_ren_em_current}
\end{align}

\section{The renormalised hadronic vacuum polarisation function in QCD+QED}

\begin{figure}
\centering
\includegraphics[width=0.49\textwidth]{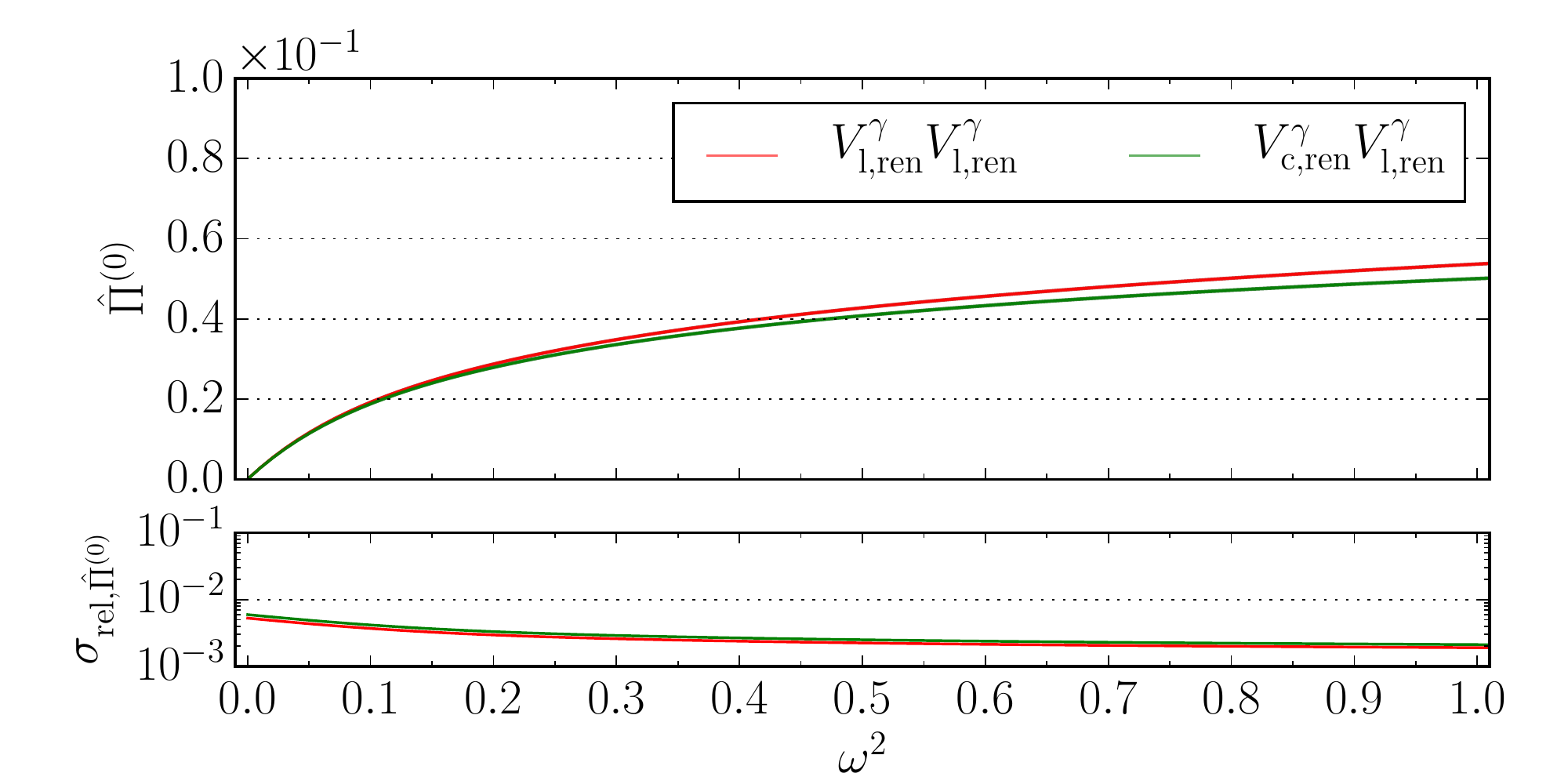}
\includegraphics[width=0.49\textwidth]{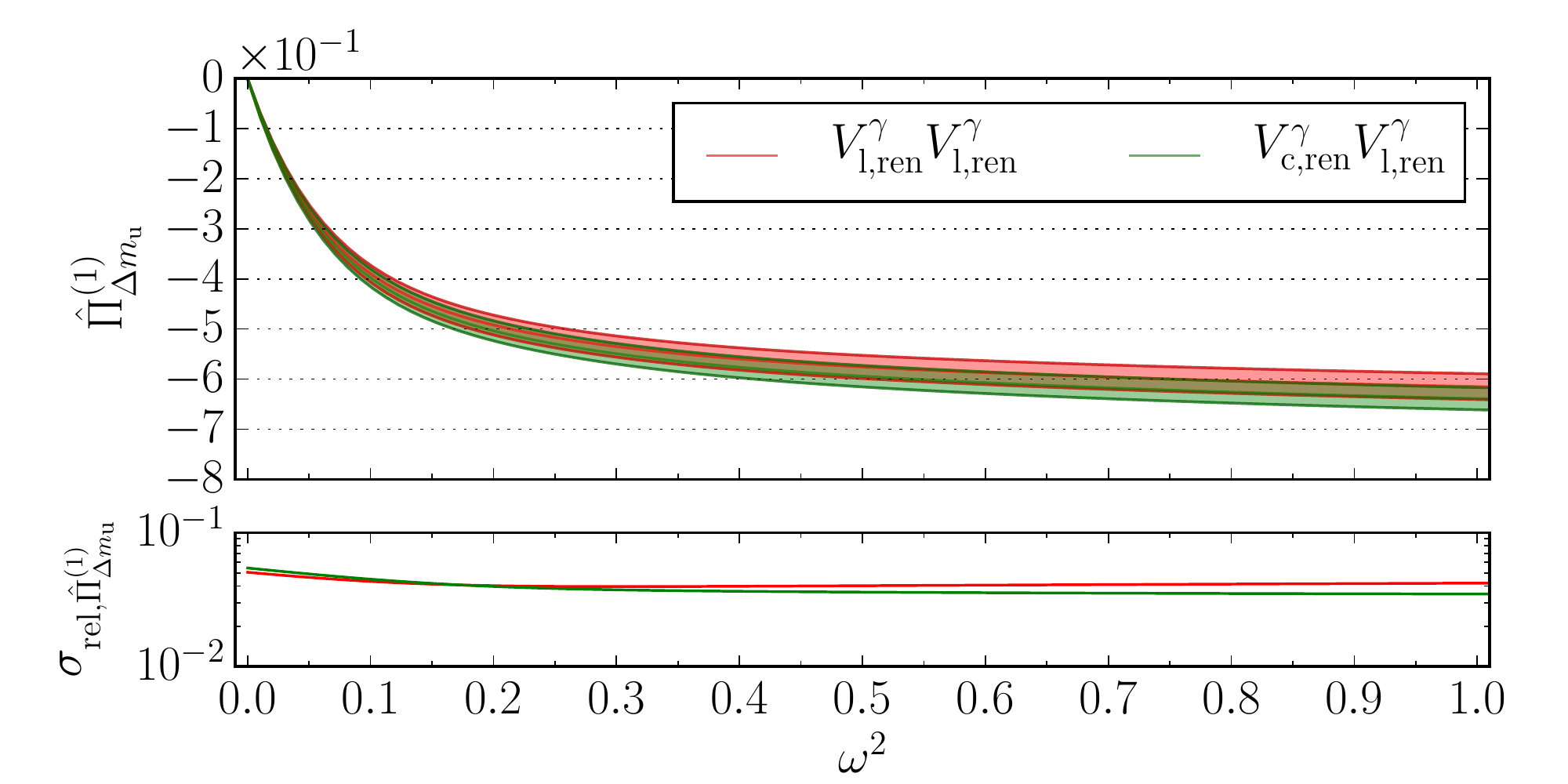}
\includegraphics[width=0.49\textwidth]{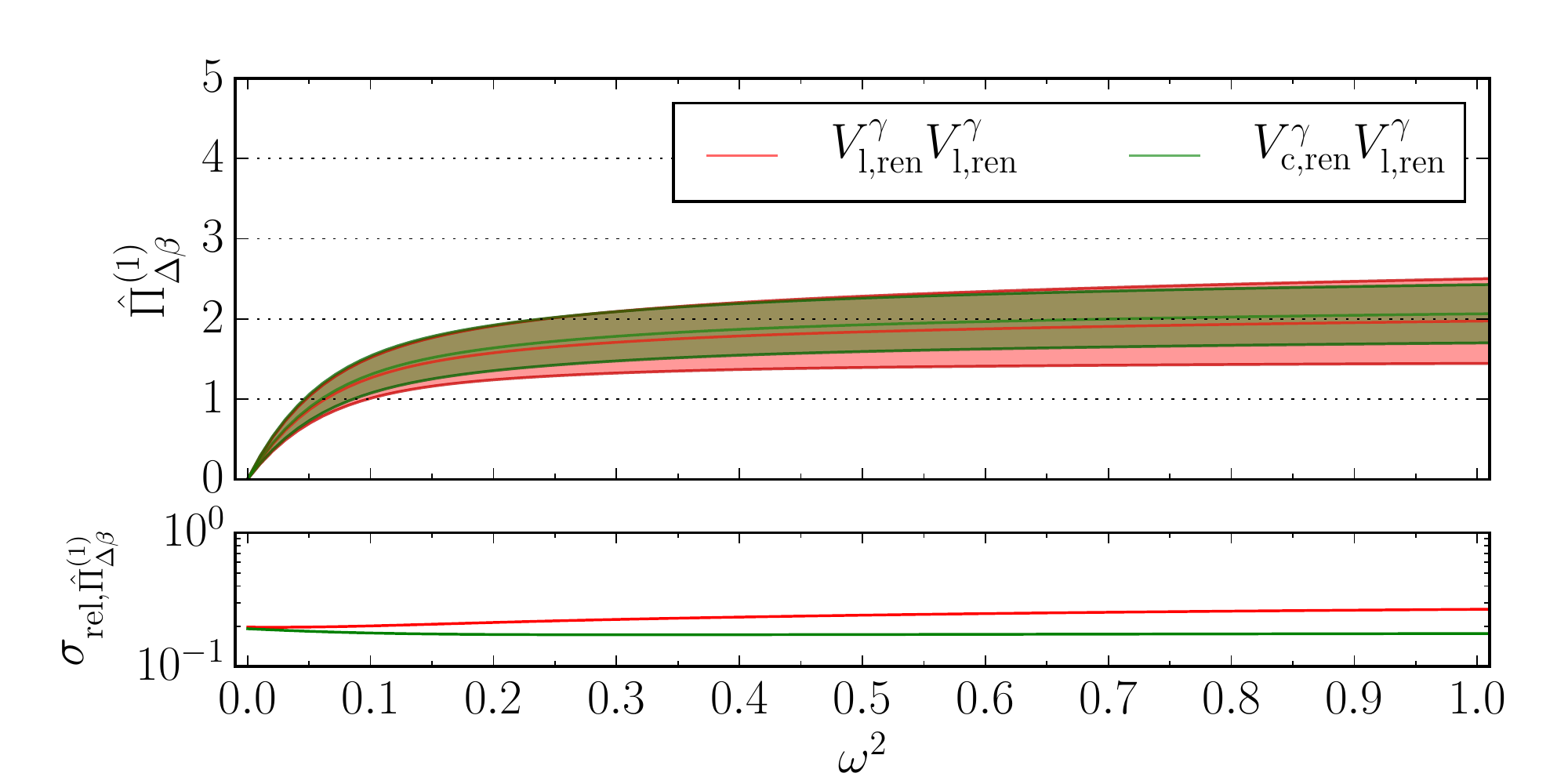}
\includegraphics[width=0.49\textwidth]{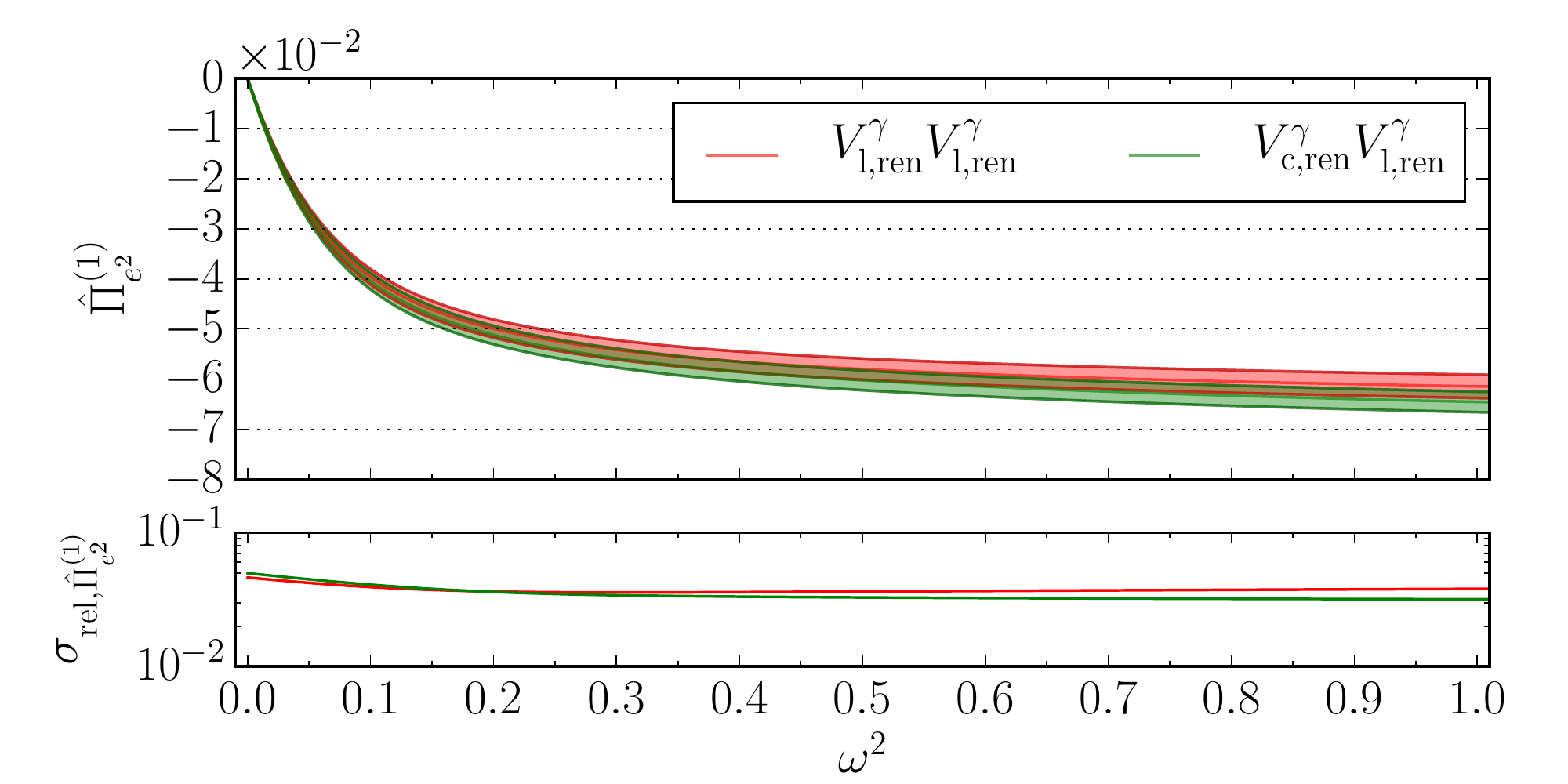}
\caption[]{Renormalised HVP function according to Eq.~\eqref{eq_hvp_function} in lattice units.}
\label{fig_hvp_function}
\end{figure}

The subtracted HVP function $\hat{\Pi}(\omega^2)=\Pi(\omega^2)-\Pi(0)$~\cite{Bernecker:2011gh} in continuum and infinite-volume QCD+QED is defined as $\hat{\Pi}(\omega^2)\delta^{\mu_{2}\mu_{1}} = \int_{0}^{\infty} \mathrm{d}x^{0}\, K(\omega^{2},x^{0})\int \mathrm{d}x^{3}\langle V^{\gamma x\mu_{2}} V^{\gamma 0\mu_{1}}\rangle_{\mathrm{QCD-con}}$ 
with the integration kernel $K(\omega^{2},t) = -\frac{1}{\omega^2}(\omega^{2}t^{2}-4\sin^{2}(\frac{\omega t}{2}))$. Making use of the renormalised electromagnetic vector-vector correlation function obtained from Eqs.~\eqref{eq_vv_corfunction} and~\eqref{eq_ren_em_current}
\begin{align*}
C^{\gamma\gamma}_{d_{2}\mathrm{l},\mathrm{ren}} &= \sum_{i_{2},i_{1}}\Big(Z_{V_{d_{2},\mathrm{ren}}^{3}V_{d_{2}}^{i_{2}}} + \frac{1}{\sqrt{3}} Z_{V_{d_{2},\mathrm{ren}}^{8}V_{d_{2}}^{i_{2}}}\Big) \Big(Z_{V_{\mathrm{l},\mathrm{ren}}^{3}V_{\mathrm{l}}^{i_{1}}} + \frac{1}{\sqrt{3}} Z_{V_{\mathrm{l},\mathrm{ren}}^{8}V_{\mathrm{l}}^{i_{1}}}\Big) C^{i_{2}i_{1}}_{d_{2}\mathrm{l}} \quad d_{2} = \mathrm{l},\mathrm{c}.
\end{align*}
and subtracting the QCD-disconnected part~\cite{Chakraborty:2018iyb} the lattice approximation to $\hat{\Pi}(\omega^2)$ reads
\begin{align}
(\hat{\Pi}_{d_{2}\mathrm{l},\mathrm{ren}}(\omega^2))^{(0)} &= \sum_{x^{0}_{2}=x^{0}_{1}}^{x^{0}_{\mathrm{cut}}} K(\omega^{2},x^{0}_{2}-x^{0}_{1}) (C{}^{\gamma\gamma}_{d_{2}\mathrm{l},\mathrm{ren}}(x^{0}_{2},x^{0}_{1}))^{(0)}, \nonumber\\
(\hat{\Pi}_{d_{2}\mathrm{l},\mathrm{ren}}(\omega^2))^{(1)}_{l} &= \sum_{x^{0}_{2}=x^{0}_{1}}^{x^{0}_{\mathrm{cut}}} K(\omega^{2},x^{0}_{2}-x^{0}_{1}) (C{}^{\gamma\gamma}_{d_{2}\mathrm{l},\mathrm{ren}}(x^{0}_{2},x^{0}_{1}))^{(1)}_{l} \quad l=\Delta m_{\mathrm{u}},\Delta m_{\mathrm{d}},\Delta m_{\mathrm{s}},\Delta\beta, \nonumber \\
(\hat{\Pi}_{d_{2}\mathrm{l},\mathrm{ren}}(\omega^2))^{(1)}_{e^{2}} &=\sum_{x^{0}_{2}=x^{0}_{1}}^{x^{0}_{\mathrm{cut}}} K(\omega^{2},x^{0}_{2}-x^{0}_{1})(C{}^{\gamma\gamma}_{d_{2}\mathrm{l},\mathrm{ren}}(x^{0}_{2},x^{0}_{1}))^{(1)}_{e^{2}} - \Big((\Pi(\omega^2))^{(0)}\Big)^{2} \label{eq_hvp_function}.
\end{align}
We choose $x^{0}_{\mathrm{cut}}=x_{1}^{0}+22$, above which the signal is lost. At larger times one may model $C^{\gamma\gamma}_{d_{2}\mathrm{l},\mathrm{ren}}$ by a truncated spectral decomposition with parameters obtained from a fit at earlier times~\cite{Gerardin:2019rua}. Fiq.~\ref{fig_hvp_function} shows results for the quark-connected contributions in Eq.~\eqref{eq_ibvalence_diagrams} to the HVP function.

\section{Quark-disconnected contributions}

For the evaluation of quark-disconnected diagrams in Eq.~\eqref{eq_ibvalence_diagrams} we used the following setup: For the quark loop at the mesonic source we used the inversions obtained from the quark-connected calculation. The quark loop at the mesonic sink was estimated using stochastic volume sources in combination with hierarchical probing~\cite{Stathopoulos:2013aci}. We used 64 Hadamard vectors and one source per configuration. At the given level of statistics we were only able to observe a signal in the pseudo-scalar channel, but not in the vector channel relevant for the HVP.

In~\cite{Giusti:2019kff} a method to reduce the noise in the calculation of quark loops based on the one-end-trick and frequency splitting in terms of quark masses was introduced. Below we discuss how this technique can be adapted in order to include isospin-breaking effects. The key insight is the fact that a difference of propagators of non-degenerate quark flavours $f_{1}$ and $f_{2}$ can be transformed into a product of propagators using $(S^{(f)})^{-1}=D^{(f)}$, where $D^{(f)}$ is the isosymmetric Dirac operator, i.e. 
\begin{align}
S^{(f_{2})}-S^{(f_{1})} = S^{(f_{2})}(D^{(f_{1})}-D^{(f_{2})})S^{(f_{1})} = -(m^{(f_{2})}-m^{(f_{1})})S^{(f_{2})}S^{(f_{1})}.\label{eq_difference}
\end{align}
For stochastic quark sources $\eta$ with $\langle\eta \eta^{\dagger}\rangle_{\eta} = \mathds{1}$ the split-even random-noise estimator reads
\begin{align}
\mathrm{Tr_{\mathrm{sc}}}(\Gamma S^{(f_{2})}{}^{xx})-\mathrm{Tr_{\mathrm{sc}}}(\Gamma S^{(f_{1})}{}^{xx}) &= -(m^{(f_{2})}-m^{(f_{1})})\Big\langle\mathrm{Tr_{\mathrm{sc}}}(\Gamma (S^{(f_{2})}\eta){}^{x} (\eta^{\dagger}S^{(f_{1})}){}^{x})\Big\rangle_{\eta}\label{eq_estimator}.
\end{align}
Introducing additional quark flavours with $m^{(f_{1})}<\ldots<m^{(f_{n})}$ one defines a frequency-splitting random noise estimator $\mathrm{Tr_{\mathrm{sc}}}(\Gamma S^{(f_{1})}{}^{xx}) = \sum_{i=1}^{n-1}\Big(\mathrm{Tr_{\mathrm{sc}}}(\Gamma S^{(f_{i})}{}^{xx}) - \mathrm{Tr_{\mathrm{sc}}}(\Gamma S^{(f_{i+1})}{}^{xx})\Big) + \mathrm{Tr_{\mathrm{sc}}}(\Gamma S^{(f_{n})}{}^{xx})$ , where $\mathrm{Tr_{\mathrm{sc}}}(\Gamma S^{(f_{i})}{}^{xx}) - \mathrm{Tr_{\mathrm{sc}}}(\Gamma S^{(f_{i+1})}{}^{xx})$ is calculated using Eq.~\eqref{eq_estimator} and $\mathrm{Tr}(\Gamma S^{(f_{n})}{}^{xx})$ is evaluated for the heaviest quark flavour $f_{n}$ via a hopping parameter expansion or hierarchical probing~\cite{Stathopoulos:2013aci}. Making use of the fact that the vertices derived from the perturbative expansion~\cite{Risch:2018ozp} are flavour diagonal and up to a normalisation factor flavour blind, i.e. $V_{\mathrm{qq}}{{}^{f_{2}}}_{f_{1}} = \delta{}^{f_{2}}_{f_{1}}\hat{V}_{\mathrm{qq}}$, $V_{\mathrm{qq}\gamma}{}^{f_{2}}{}_{f_{1}} = \delta{}^{f_{2}}_{f_{1}}e^{(f_{1})}\hat{V}_{\mathrm{qq}\gamma}$ and $V_{\mathrm{qq}\gamma\gamma}{}^{f_{2}}{}_{f_{1}} = \delta{}^{f_{2}}_{f_{1}}(e^{(f_{1})})^{2}\hat{V}_{\mathrm{qq}\gamma\gamma}$, we can extend the latter approach to quark-disconnected contributions in the isospin-breaking expansion. The difference of sequential propagators over the vertices $V=\hat{V}_{\mathrm{qq}}, \hat{V}_{\mathrm{qq\gamma}}{}_{\mathbf{c}}A{}^{\mathbf{c}}, \hat{V}_{\mathrm{qq\gamma\gamma}}{}_{\mathbf{c_{2}}\mathbf{c_{1}}}A_{2}{}^{\mathbf{c_{2}}}A_{1}{}^{\mathbf{c_{1}}}$ contracted with stochastic photon fields $A$, where $\mathbf{c}\equiv x\mu$, can also be written as a sum of products of a propagator and a sequential propagator making use of Eq.~\eqref{eq_difference}:
\begin{align*}
S^{(f_{2})}VS^{(f_{2})}-S^{(f_{1})}VS^{(f_{1})} &= S^{(f_{2})}V(S^{(f_{2})}-S^{(f_{1})})+(S^{(f_{2})}-S^{(f_{1})})VS^{(f_{1})} \\
&= -(m^{(f_{2})}-m^{(f_{1})})(S^{(f_{2})}VS^{(f_{2})}S^{(f_{1})}+S^{(f_{2})}S^{(f_{1})}VS^{(f_{1})}).
\end{align*}
The split-even random-noise estimator for quark loops including a vertex then reads
\begin{align*}
& \mathrm{Tr}_{\mathrm{sc}}(\Gamma (S^{(f_{2})}VS^{(f_{2})}){}^{xx})-\mathrm{Tr}_{\mathrm{sc}}(\Gamma (S^{(f_{1})}VS^{(f_{1})}){}^{xx}) \\
&= -(m^{(f_{2})}-m^{(f_{1})})\Big\langle\mathrm{Tr}_{\mathrm{sc}}(\Gamma (S^{(f_{2})}VS^{(f_{2})}\eta){}^{x} (\eta^{\dagger}S^{(f_{1})}){}^{x})+\mathrm{Tr}_{\mathrm{sc}}(\Gamma (S^{(f_{2})}\eta){}^{x} (\eta^{\dagger}S^{(f_{1})}VS^{(f_{1})}){}^{x})\Big\rangle_{\eta}.
\end{align*}
Double sequential propagators $S^{(f)}V_{2}S^{(f)}V_{1}S^{(f)}$ are treated in a similar fashion.

\section{Conclusions and Outlook}

In order to construct the renormalised HVP function in QCD+QED we introduced the perturbative expansion of the vector-vector correlation functions for leading isospin-breaking effects and determined the renormalisation factors for the local vector current including operator mixing on one ensemble. We also commented on the status of our investigation of quark-disconnected contributions, introducing a suitable adaption of the one-end trick to perturbative isospin-breaking. We plan to extend our analysis by the hadronic contribution to the anomalous moment of the muon, where we reconstruct the large-time behaviour of the vector-vector correlation function by the ground state mass, as well as by baryon masses used for the scale setting.
\vspace{0.5em}

\begin{small}
We are grateful to our colleagues within the CLS initiative for sharing ensembles. Our calculations were performed on the HPC Cluster "Clover" at the Helmholtz Institute Mainz and on the HPC Cluster "Mogon II" at the University of Mainz.
\end{small}

\bibliographystyle{JHEP}
\bibliography{proceedings.bib}

\end{document}